\documentclass[prx,superscriptaddress,aps,amsmath,amssymb,floatfix,reprint,raggedbottom,longbibliography]{revtex4-2}
\usepackage{graphicx}
\usepackage{balance}
\usepackage{hyperref}
\hypersetup{
    colorlinks=true,       
    linkcolor=red,          
  citecolor=magenta,        
    filecolor=magenta,      
    urlcolor=red,           
    runcolor=cyan
}
\usepackage{amsmath}
\usepackage{times}

\usepackage{dcolumn}
\usepackage{xcolor}
\usepackage{lipsum}
\usepackage{soul}
\usepackage{braket}
\usepackage{amsfonts}
\usepackage{multirow}
\usepackage[utf8]{inputenc}
\DeclareUnicodeCharacter{2009}{\,}

\newcommand{\beginsupplement}{%
    \setcounter{table}{0}	
    \renewcommand{\thetable}{S\arabic{table}}%
    \setcounter{figure}{0}	
    \renewcommand{\thefigure}{S\arabic{figure}}%
    \setcounter{equation}{0}	
    \renewcommand{\theequation}{S.\arabic{equation}}%
    \renewcommand{\thesection}{}%
    \setcounter{subsection}{0}
    \renewcommand\thesubsection{\Alph{subsection}}%
 }

\makeatletter 
\renewcommand{\fnum@figure}{\textbf{Fig.~\thefigure}}
\makeatother

\makeatletter
\def\bbordermatrix#1{\begingroup \m@th
  \@tempdima 4.75\p@
  \setbox\z@\vbox{%
    \def\cr{\crcr\noalign{\kern2\p@\global\let\cr\endline}}%
    \ialign{$##$\hfil\kern2\p@\kern\@tempdima&\thinspace\hfil$##$\hfil
      &&\quad\hfil$##$\hfil\crcr
      \omit\strut\hfil\crcr\noalign{\kern-\baselineskip}%
      #1\crcr\omit\strut\cr}}%
  \setbox\tw@\vbox{\unvcopy\z@\global\setbox\@ne\lastbox}%
  \setbox\tw@\hbox{\unhbox\@ne\unskip\global\setbox\@ne\lastbox}%
  \setbox\tw@\hbox{$\kern\wd\@ne\kern-\@tempdima\left[\kern-\wd\@ne
    \global\setbox\@ne\vbox{\box\@ne\kern2\p@}%
    \vcenter{\kern-\ht\@ne\unvbox\z@\kern-\baselineskip}\,\right]$}%
  \null\;\vbox{\kern\ht\@ne\box\tw@}\endgroup}
\makeatother

\emergencystretch=\maxdimen
\usepackage[compact]{titlesec}
\titlespacing{\section}{0pt}{*3}{*2}
\titlespacing{\subsection}{0pt}{*2}{*2}
\titlespacing{\subsubsection}{0pt}{*2}{*2}

\titleformat{\section}{\filcenter\normalfont\small \bfseries}{\thesection.}{1em}{\MakeUppercase}   

\begin{document}
\title{Massively Parallel Probabilistic Computing with Sparse Ising Machines}

\author{Navid Anjum Aadit}\email{maadit@ece.ucsb.edu}
\affiliation{Department of Electrical and Computer Engineering, University of California, Santa Barbara, Santa Barbara, CA, 93106, USA}
\author{Andrea Grimaldi}
\affiliation {Department of Mathematical and Computer Sciences, Physical Sciences and Earth Sciences, \\University of Messina, Messina, Italy}
\author{Mario Carpentieri}
\affiliation{Department of Electrical and Information Engineering, Politecnico di Bari, Bari, Italy}
\author{Luke Theogarajan}
\affiliation{Department of Electrical and Computer Engineering, University of California, Santa Barbara, Santa Barbara, CA, 93106, USA}
\author{John M. Martinis}
\affiliation{Department of Physics, University of California, Santa Barbara, Santa Barbara, CA, 93106, USA}
\affiliation{Quantala, Santa Barbara, CA, 93105, CA, USA}

\affiliation{Zyphra Technologies Inc., San Francisco, CA, USA}
\author{Giovanni Finocchio}\email{giovanni.finocchio@unime.it }
\affiliation {Department of Mathematical and Computer Sciences, Physical Sciences and Earth Sciences, \\University of Messina, Messina, Italy}
\author{Kerem Y. Camsari}\email{camsari@ece.ucsb.edu}
\affiliation{Department of Electrical and Computer Engineering, University of California, Santa Barbara, Santa Barbara, CA, 93106, USA}

\date{\today}
\begin{abstract}

Inspired by the developments in quantum computing, building domain-specific classical hardware to solve computationally hard problems has received increasing attention. 
Here, by introducing systematic sparsification techniques, we demonstrate a massively parallel architecture: the sparse Ising Machine (sIM). Exploiting sparsity, sIM  achieves ideal parallelism: its key figure of merit $-$ flips per second $-$ scales linearly with the number of  probabilistic bits (p-bit) in the system. This makes sIM  up to 6 orders of magnitude faster than a CPU implementing standard Gibbs sampling. Compared to optimized implementations in TPUs and GPUs, sIM  delivers 5-18x speedup in sampling. In benchmark problems such as integer factorization, sIM can reliably factor semiprimes up to 32-bits, far larger than previous attempts from D-Wave and other probabilistic solvers. Strikingly, sIM  beats competition-winning SAT solvers (by 4-700x in runtime to reach 95\% accuracy) in solving 3SAT problems. Even when sampling is made inexact using faster clocks, sIM can find the correct ground state with further speedup. The problem encoding and sparsification techniques we introduce can be applied to other Ising Machines (classical and quantum) and the architecture we present can be used for scaling  the demonstrated 5,000$-$10,000 p-bits to 1,000,000 or more  through analog CMOS or nanodevices.

\end{abstract}
 \maketitle

\section{Introduction}
\label{sec:Intro}
 Markov Chain Monte Carlo (MCMC) algorithms have made  a significant impact  in the history of computing \cite{metropolis1953equation}. MCMC methods are among the most powerful randomized algorithms with a wide range of applications in Artificial Intelligence (AI) \cite{buluc2021randomized}. Powerful MCMC methods such as Metropolis and Gibbs sampling have been widely applied to training generative neural networks \cite{hinton2012practical}, probabilistic inference  in belief networks \cite{mansinghka2008stochastic}, calculating physical observables in classical and quantum systems \cite{bouchard2018bouncy,krauth1996quantum} and solving computationally hard combinatorial optimization problems \cite{kirkpatrick1987optimization}.

Designing domain-specific hardware to accelerate such computationally hard problems of AI has been receiving increasing attention with the slowing pace of Moore's Law. There have been a number of approaches to build special-purpose hardware to solve computationally hard problems. A class of such solvers (also known as Ising Machines) specifically solve quadratic energy models or the Ising model, typically mapped to problems in NP \cite{mcmahon2016fully,Hitachi_yamaoka2015,Toshiba_goto2019combinatorial,wang2019_oim,ahmed2020probabilisticosc,chou2019analog,dutta2021ising,borders2019integer,Janus_baity2014janus,Fujitsu_aramon2019physics,STATICA_yamamoto2020,patel2020ising,patel2020logically,su2021252,smithson2019efficient,cai2020power,tatsumura2021scaling,sutton2020autonomous},
\begin{equation}
     E = - \left(\sum_{i<j} J_{ij} m_i m_j + \sum h_i m_i \right) \vspace{-4pt}
     \label{eq:en}
\end{equation}
($m_i$ $\in$ $\pm$ 1, $J_{ij}$=$J_{ji}$, $h_i$ $\in$ $\mathbb{R}$), where quadratic terms ($m_i m_j$) in the energy translate to a linear ``synapse'' or an interconnection matrix which can be expressed as a graph ($I_i=-\partial E / \partial m_i $):
\begin{equation}
    I_i = \sum J_{ij} m_j + h_i 
    \label{eq:syn}
\end{equation}
Choosing the activation function of individual probabilistic bits as
\begin{equation}
 m_i = \mathrm{sgn}\left[\mathrm{tanh}(\beta I_i)- \mathrm{rand}_U(-1,1)\right]
 \label{eq:pbit}
\end{equation}
ensures the system states, $\{m\}$, are visited according to their corresponding Boltzmann probability:
\begin{equation}
    p_i \propto \mathrm{exp}\left[-\beta E(\{m\})\right]
    \label{eq:boltz} 
\end{equation}
where $\beta$ is introduced as an ``inverse temperature'' and can be used to enhance or suppress  probabilities corresponding to energy minima.  The dynamical evolution of Eq.~\eqref{eq:syn}-\eqref{eq:pbit} enables probabilistic sampling and inference, learning weights of a stochastic neural network or performing search or optimization in the exponential state space of the model. Such a machine evolving to the Boltzmann distribution defined by Eq.~\eqref{eq:boltz} is called a Boltzmann Machine, after the pioneering work of Hinton and colleagues \cite{ackley1985learning,hinton2007boltzmann}.

So far, nearly all dedicated Ising Machines have specifically focused on optimization problems, with the exception of D-Wave's quantum annealers which have been applied to problems beyond combinatorial optimization \cite{dixit2021training,koshka2020toward}. 

Typically in Gibbs sampling (a type of MCMC method),  Eq.~\eqref{eq:syn}-\eqref{eq:pbit} are updated iteratively to dynamically evolve the Markov chain such that the network  eventually reaches the Boltzmann distribution defined by Eq.~\eqref{eq:boltz}. A practical difficulty lies in the serial nature of this evolution: connected nodes  need to be updated one after the other since parallel updating leads to repeated oscillations in the network state, preventing the network from converging to the Boltzmann distribution. The need for sequential updating inherently serializes the network evolution, signified by the nested for-loops in standard descriptions of Gibbs sampling \cite{koller2009probabilistic}. 

\section{Summary of Main Results} 
\label{sec:summary_results}
This work is about overcoming the fundamental difficulty of sequential updating by combining algorithmic and architectural ideas to build a sparse Ising Machine (sIM) for solving combinatorial optimization and probabilistic sampling problems.  The present implementation is on a Field Programmable Gate Array (FPGA)  however the architecture is general and can have many different implementations ranging from digital CMOS to energy-efficient  nanodevices such as Magnetic Tunnel Junctions \cite{finocchio2021promise, grollier2020neuromorphic} (see Supplementary Information  Section~\ref{sec:characteristic_devices}, ~\ref{sec:fpga_implement}).

The sIM  achieves near-ideal parallelism for MCMC sampling as long as the  interconnection matrix $J$ is sparse. A key feature of our framework is in its generality: we first show that \emph{any} optimization function can be efficiently represented as a sparse (but irregular) graph through principles of invertible logic \cite{camsari2017stochasticL,smithson2019efficient}. We then outline techniques for further sparsification using additional nodes without any approximation. Next, we develop a massively parallel architecture to implement Eq.~\eqref{eq:syn}-\eqref{eq:pbit}, exploiting the sparsity of the graph, where the graph is defined by $J$. This is achieved by using multiple phase shifted clocks
controlling the activation of probabilistic bits (p-bits) (Eq.~\eqref{eq:pbit}). The p-bits are interconnected through a  multiply-accumulate (MAC) unit (Eq.~\eqref{eq:syn}).

This architecture can be considered to be a low level hardware-level implementation of chromatic Gibbs sampling \cite{gonzalez2011parallel} where large blocks of conditionally independent nodes are updated in parallel. For this sampling to be exact, the MAC must finish its computation before the next color block is updated. An unexpected finding however is even when color blocks are updated before the MAC operation is completed, the network is often able to find exact ground states in model optimization problems. This inexact Gibbs sampling approach is reminiscent of the Hogwild!-Gibbs algorithm \cite{johnson2013analyzing} and we show how this overclocking strategy can lead to further advantages. We provide error models and an analysis of inexact Gibbs sampling with analytical estimates of limiting behaviors.

The idea of block updating is commonly used for regular graphs. For example, as  first noted in Ref.~\cite{geman1984stochastic}, when the graph is bipartite (as in Restricted Boltzmann Machines or chessboard lattices), trivial colorings (with two colors, black and white or four colors in King's graphs) are possible and this is often exploited in updating each color block in parallel \cite{ko2019flexgibbs,patel2020ising,mansinghka2008stochastic,fang2014parallel,yang2019high,yoshimura2016fpga,yoshimura2017implementation}. We also note that parallelization techniques in multiple FPGAs have recently been explored for other types of Ising Machines \cite{tatsumura2021scaling}, however these are based on completely different algorithms unrelated to the computational model we use based on Eq.~\eqref{eq:syn}-\eqref{eq:boltz}.

Compared to prior works on block updating \cite{ko2019flexgibbs,patel2020ising,mansinghka2008stochastic,fang2014parallel,yang2019high,yoshimura2016fpga,yoshimura2017implementation}, our contributions are twofold: 
First, we extend the block updating scheme such that it applies to regular \emph{and} irregular graphs with the only requirement that the graphs are sparse enough to be colored by a few colors (typically $\le$ 4-8).  This generalization is significant, considering most practical instances of combinatorial optimization problems have irregular graph representations. 

Second, we provide exact sparsification methods which can be applied to sparsify dense graphs. We believe that both of these methods can be useful for other Ising Machines and different problems. 

We tested the resultant sIM  on model problems and achieved three key results: 
\begin{itemize}
    \item In solving  Boolean satisfiability problems, the sIM  is able to beat competition-winning SAT solvers (2020, 2017) in run-time by up to 4-700x to reach 95\% accuracy.
    \item In probabilistic sampling, the sIM  delivers a measured 5-18x speedup over the optimized TPU and GPU implementations (Table~\ref{tab:benchmarking}). Against a standard CPU implementation, we measure up to 6-orders of magnitude speedup. 
    \item In integer factorization, the sIM  can reliably find the absolute ground state for semiprimes up to 32-bits, far larger than what has been reported for D-Wave's quantum annealers or similar probabilistic solvers \cite{andriyash2016boosting, borders2019integer, smithson2019efficient,dridi2017prime, jiang2018quantum, patel2020logically} (Table~\ref{tab:factors}). Robust factorization up to 32-bit numbers seem to be the largest by far among these alternatives.  \end{itemize}
In this paper, we use integer factorization as a computationally hard optimization problem to compare the performance of sIM with respect to D-Wave and other Ising Machines. Expressing the integer factorization problem as a satisfiability (or spin-glass) instance is not expected to be practically relevant, as studied in detail in Ref.~\cite{mosca2019factoring}. However, critical subroutines of the best algorithms for factoring may potentially be accelerated using improved satisfiability through dedicated hardware or algorithms \cite{mosca2020speeding}.

A striking result is to show speedups over recent competition-winning SAT solvers in approximate optimization,  since SAT solvers have been optimized and fine-tuned after decades of research and development. In contrast, the sIM  is using a standard simulated annealing algorithm without any detailed fine-tuning. Further improvements using more sophisticated algorithms such as Parallel Tempering should increase the performance of the sIM. Moreover, experimental developments in emerging nanodevice technologies \cite{safranski2021demonstration,hayakawa2021nanosecond} such as Magnetic Tunnel Junction based asynchronous probabilistic computers \cite{borders2019integer} can use the same architectural and algorithmic ideas we develop in this paper to provide additional speedup \cite{sutton2020autonomous}.

The organization of this paper is as follows: Section~\ref{sec:composition_reconfig} shows how any combinatorial optimization problem can be converted to an invertible  probabilistic circuit (p-circuit) using the basic building blocks (AND, OR, NOT, Full Adder) with compact interconnection matrices and discrete weights, followed by a discussion on reconfigurability. We illustrate in Section~\ref{sec: architecture_graph_coloring} how a given sparse p-circuit can be colored with a few colors using an approximate  graph coloring algorithm, DSATUR \cite{brelaz1979new}.  We elaborate on the details of the hardware architecture implementing the coupled equations and propose further sparsification techniques to overcome the limitations of high fan-out circuits. Next, we report our experimental results on integer factorization and Boolean satisfiability problems  in Sections~\ref{sec: Results_fact}-\ref{sec:SAT} as well as detailed comparisons of the sIM  with respect to CPU, GPU and TPU implementations of MCMC. Finally, in Section~\ref{sec: overclocking} we discuss the effects of overclocking p-bits to perform inexact and asynchronous Gibbs sampling for further improvement. 

\begin{figure*}[t!]
    \centering
    \includegraphics[width =1 \textwidth ]{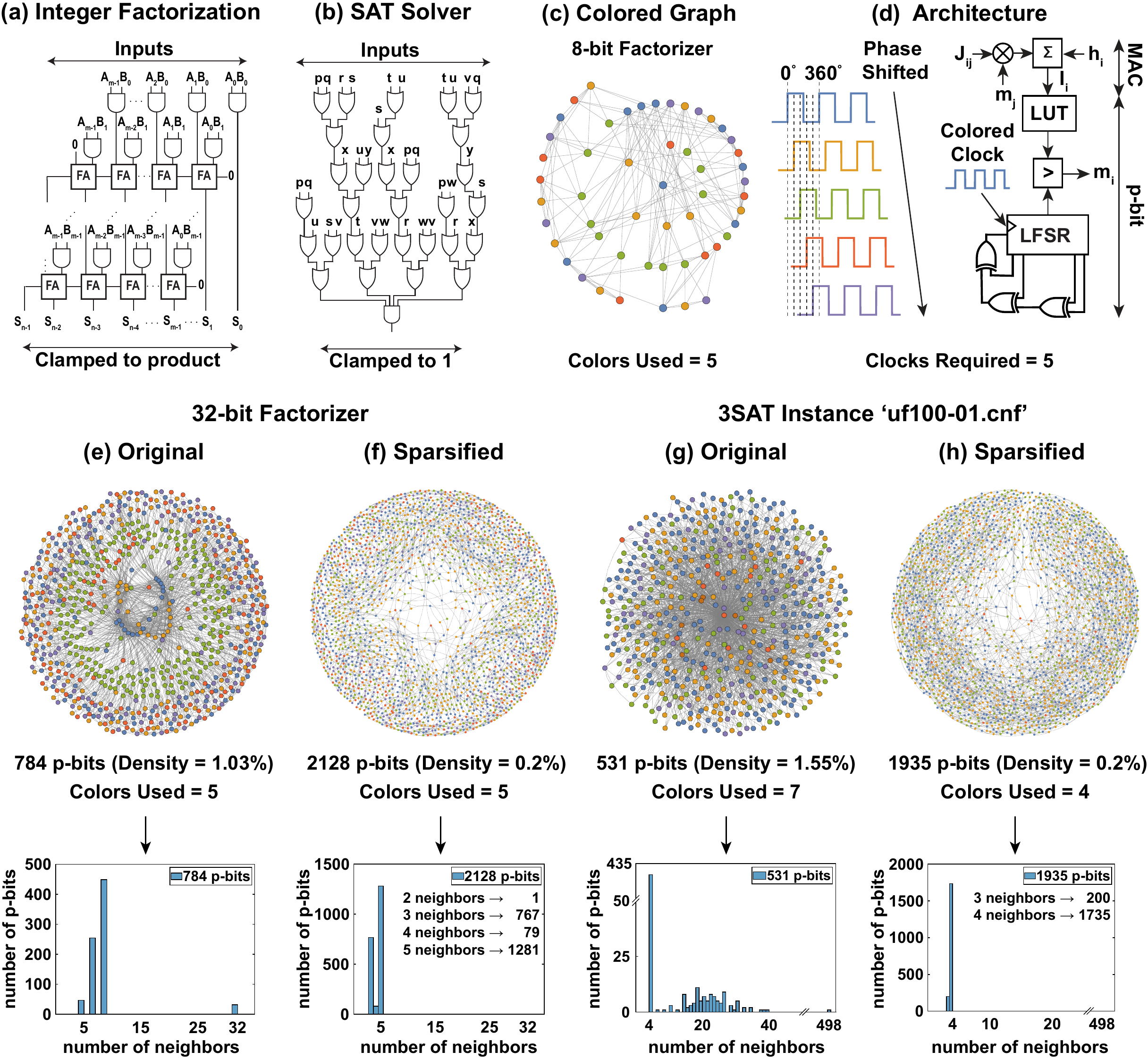}
    \caption{Combinatorial optimization with invertible logic. (a) A digital multiplier circuit composed of AND gates and full adders (FAs) that can be run invertibly to solve $n$-bit integer factorization problem. (b) A Boolean circuit composed of OR and AND gates that can be run invertibly to solve satisfiability problems. (c) An example 8-bit factorizer graph colored with 5 colors. (d) An example architecture with 5 parallel and equally phase-shifted clocks, implementing the MAC (Eq.~\eqref{eq:syn}) and the p-bit (Eq.~\eqref{eq:pbit}) equations. (e) The original graph of a 32-bit factorizer developed using invertible logic requires 5 colors but it contains nodes with up to 32 neighbors, limiting hardware implementations. (f) Exact sparsification techniques are used to reduce nodes with a large number of neighbors at the expense of additional bits. (g) The original version of the 3SAT instance `uf100-01.cnf' developed using invertible logic. (h) The sparsified version with nodes having a maximum of 4 neighbors only. Graph density, $\rho$ is defined by Eq.~\eqref{eq:rho}.} 
    \label{fig:fig1}
\end{figure*}
        
\section{Composability and Reconfigurability}
\label{sec:composition_reconfig}
\subsection{Composing invertible circuits from elementary gates}
\label{sec:composition}

Invertible logic (discussed in detail in Refs.~\cite{camsari2017stochasticL,onizawa2019design}) allows composing probabilistic generalizations of a universal set of basic gates such as COPY/NOT (2 p-bits), AND (3 p-bits), OR (3 p-bits) and Full Adders (FA, 5 p-bits). The basics of such invertible gates and their operation have been discussed extensively \cite{biamonte2008nonperturbative,pervaiz2017hardware, smithson2019efficient}. Here, we show some of the  $J$ matrices for a set of elementary probabilistic gates in the  Supplementary Information Section~\ref{sec:basic} and \ref{sec:fs}, where we also  elaborate on how to compose p-circuits corresponding to any given Boolean function.  Similar to conventional digital circuits, the use of basic p-logic gates to a hierarchical design of larger p-circuits results in highly sparse representations amenable to the massive parallelism discussed in this paper. 

The ability of invertible logic circuits to operate in reverse gives rise to a convenient method of solving inverse problems, as noted in the related paradigm of memcomputing and by D-Wave \cite{traversa2017polynomial,andriyash2016boosting}. Therefore, hard combinatorial optimization problems such as integer factorization and Boolean satisfiability (SAT) can be solved in hardware using invertible multipliers or by invertible Boolean circuits corresponding to a given satisfiability instance.  Fig.~\ref{fig:fig1}a shows a classic $m$-bit multiplier circuit composed of multiple AND gates and FAs. In the reverse direction, this circuit works as an $n$-bit factorizer circuit ($n = 2m$) where we clamp the output bits to the $n$-bit product to be factored. Similarly, we clamp the output of the SAT solver circuit in Fig.~\ref{fig:fig1}b to 1 and find the input variables satisfying all the clauses in the reverse direction. Each clause is represented by multiple OR gates; if any input variable is negated, we modify the corresponding weight ($J$) matrix of the OR gate accordingly. The output bits of the OR gates can be clamped to 1 either directly or by using an AND gate at the end. In this work, we also focus on the 3SAT, a special form of the satisfiability problem where each clause has exactly three variables in the conjunctive normal form (CNF) \cite{arora2009computational}. We collected 3SAT instances as .CNF files from the UBC SATLIB library \cite{satlib} and map them to  interconnection matrices ($J$), composed of  elementary probabilistic gates. Further details about instances used in this work are provided in the Methods Section~\ref{sec:prob_descript}. 

\subsection{Reconfigurability}
\label{sec:reconfiguability}
The invertible circuits composed out of elementary probabilistic gates are reconfigurable, able to accommodate different instances of a given problem. For example, a 32-bit factorizer can factor any two numbers of up to 32-bits by a suitable clamping of the bias values ($h_i$) of the output bits. Similarly, invertible Boolean circuits can be designed to accommodate many different instances of the SAT problem to function as general SAT solvers. For example, a 250-variable SAT solver can solve a 50-variable SAT instance by an appropriate clamping and multiplexing of input bits. 

The reconfigurability of invertible logic circuits provides an alternative to the usual method of embedding a native graph to a target graph referred to as the minor graph embedding (MGE) problem \cite{kaminsky2004scalable,choi2008minor}. As we show in the Supplementary Information Section~\ref{sec:MGE}, typical MGE algorithms often fail to find a mapping for the problems we considered, and even when a mapping is found the number of auxiliary spins is too large \cite{sugie2021minor}. Reconfigurability using invertible logic with sparsification \cite{kato2021design,onizawa2019design} is a much more practical alternative to MGE for the problems we considered. For example,
reconfigurable 32-bit factorization graph requires a Chimera target  with $\approx$ 10,000 spins, while the sparsification technique  introduced in this work requires only $\approx$ 2,000 spins. (see Supplementary Information Section~\ref{sec:MGE} for details). 

We also note that the universal nature of the Boolean satisfiability problem enables another layer of reconfigurability since many combinatorial optimization problems are mapped to satisfiability instances with minimal overhead \cite{biere2009handbook}. For example, the Maximum-Cut problem, a common benchmark for many Ising Machines, can be efficiently mapped to a Boolean satisfiability instance \cite{grimaldi2022spintronics}. 

\section{Architecture Design for Massive Parallelization}
\label{sec: architecture_graph_coloring}
\subsection{Graph coloring} 
\label{sec: graph_coloring}
When a quadratic energy model described by Eq.~\eqref{eq:en} is chosen, an invertible logic p-circuit can be represented as a graph where each node represents a p-bit and each edge represents the connection between the p-bits. Fig.~\ref{fig:fig1}c illustrates the graph of an 8-bit factorizer p-circuit encoded with 52 p-bits. 

As discussed in Section~\ref{sec:summary_results}, the coloring is used to exploit the trick that allows the parallel update of  unconnected (conditionally independent) p-bits. We first color the graph using a heuristic graph coloring algorithm DSATUR \cite{brelaz1979new}. Coloring a graph with exactly the minimum possible colors is NP-hard \cite{lucas2014ising}, however, this is unimportant for our purposes since we use DSATUR as a greedy algorithm  which may or may not find the optimum coloring. 

We find that when the overall graph density is low (e.g., $\lessapprox$ 1\%), irregular graphs containing nodes with hundreds of neighbors can be colored by a few colors in line with theoretical results on coloring sparse graphs \cite{ghaffari2017simple}.
For example, for the 8-bit factorizer graph only five colors are used and as a result, we need five parallel and equally phase-shifted clocks in the sIM  (Fig.~\ref{fig:fig1}d). The equal phase shift is required to avoid any concurrent edges of the clocks.  In our architecture (Fig.~\ref{fig:fig1}d), different color blocks receive different clocks to their RNGs, ensuring no neighboring p-bits flip at the same time for exact Gibbs sampling. This is a constraint we relax in Section~\ref{sec: overclocking}.

In the case of exact Gibbs sampling, our architecture ensures the entire network is updated in parallel in one clock period (of any color) while ensuring an effectively sequential operation.

As a quantitative measure of sparsity in the resultant invertible p-circuits we use the notion of graph density $\rho$. For an undirected graph $\rho$ is defined as
\begin{equation}
\rho = \displaystyle \frac{2|E|} {|V|^2-|V|}
\label{eq:rho}
\end{equation}
where $|V|$ is the number of nodes (vertices), corresponding to the number of p-bits in the sIM  and $|E|$ is the number of  edges, corresponding to the interconnections in the $[J]$ matrix. With this definition an all-to-all (or a complete) graph has a $\rho$ of 100\%. We find that when p-circuits are composed of universal gates, both the factorization and the satisfiability graphs are sparse (see Supplementary Fig.~\ref{fig:density}). For example, consider the graph of a 32-bit factorizer p-circuit presented in Fig.~\ref{fig:fig1}e having 784  p-bits with  a $\rho$ of 1.03\%, requiring only 5 distinct colors. In this example, five parallel and equally phase-shifted clocks in the sIM  are adequate to implement this p-circuit for massively parallel computation.  

\subsection{Sparsification of problem graphs} 
\label{sec: sparsification}
There is a limitation on the clock speed depending on the maximum number of neighbors for each p-bit. The neighbor distribution of the 32-bit factorizer graph reveals that it has 32 p-bits with 32 neighbors (Fig.~\ref{fig:fig1}e). In order to ensure every single color block is updated with the latest state of each neighbor, the MAC unit 
implementing  Eq.~\eqref{eq:syn} needs to finish its computation before the next color block is updated. With binary models the multiplication consists of simple multiplexing (the weights are either selected or they are ignored). This means that the addition for the weights needs to be completed within the $n^{th}$ of a clock period, where $n$ is the number of colors. In the present example, this requires large adders to add 32 $s$-bit numbers within this short period, where $s$ is the bit precision of the weights. 

Therefore even when the sparse graphs as in Fig.~\ref{fig:fig1}e,g require a few colors the p-bits with 32 or 498 neighbors introduce large synapse (addition) delays creating a severe bottleneck for how fast the clocks can be operated. To overcome this limitation, we have developed exact sparsification techniques to remove the nodes with a large number of neighbors without changing the structure of the optimization problem. The main idea in this approach is to split a given p-bit representation between two p-bits  coupled by a ferromagnetic ($J>$0) interaction which we call a COPY gate \cite{kaminsky2004scalable,choi2008minor,patel2020ising}. The ferromagnetic interaction  ensures that at the end of an annealing schedule (high $\beta$), the ground states of the split and fused models are identical. We show a formal proof in the Supplementary Information Section~\ref{sec:fs}. 

Fig.~\ref{fig:fig1}f shows the sparsified graph of the 32-bit factorizer p-circuit with 2128 p-bits and a graph density of 0.2\%. It is colored with 5 colors as before, however, the neighbor distribution reveals the maximum number of neighbors, $k$ is limited to 5 which minimizes the adder delay and allows fast clocks. In general, such sparsification techniques always introduce extra p-bits, however, in scaled implementations individual p-bits are almost always cheaper than complicated synapse interactions.

Similarly, the original graph of the 3SAT instance `uf-100-01.cnf' can be colored using 7 colors (Fig.~\ref{fig:fig1}g). It is also a sparse graph with 531 p-bits and a graph density of 1.55\%. However, the neighbor distribution shows that some p-bits have more than 5 connections and one p-bit has 498 connections. This p-bit corresponds to the node where the outputs of all clauses meet. A sparsified version of this graph is illustrated in Fig.~\ref{fig:fig1}h with 1935 p-bits and a graph density of 0.2\%. As before, the sparsification ensures the graph has a maximum of $k = 4$ neighbors for each p-bit and hence avoids large adder delays, requiring 4 colors (clocks). 

Even though we show specific cases of sparsification (with $ k = 4$ neighbors for satisfiability and $k = 5$ for factorization), we have analyzed the effect of sparsification (as a function of $k$) on system size and performance in the Supplementary Information Section~\ref{sec:scalability}. We note that limiting the number of neighbors per p-bit to a \emph{fixed} value ($ k = 4$, $k = 8$ etc.)  ensures that the adder delays from the MAC unit do not grow with system size. In other words, no matter how large the global system becomes, only the local neighborhood of a p-bit (with $k$-neighbors) needs to communicate faster than the  p-bit clocks, ensuring the scalability of the approach.

In the following sections, we present our results for integer factorization and SAT solving implemented in the massively parallel sIM  architecture. We implemented the sIM  architecture on a Xilinx Virtex UltraScale+ FPGA VCU118 Evaluation board for our experiments. The details of the FPGA architecture and its design choices are included in the Supplementary Information Section~\ref{sec:fpga_implement}. We used a simple simulated annealing algorithm \cite{aarts1989simulated} with a linear schedule for all experiments reported in this work. 

\section{Comparing Sparse Ising Machine  with existing hardware (CPU, GPU and TPU)}
\label{sec: Results_fact}

\begin{figure*}[t]
    \centering
    \includegraphics[width=1 \textwidth]{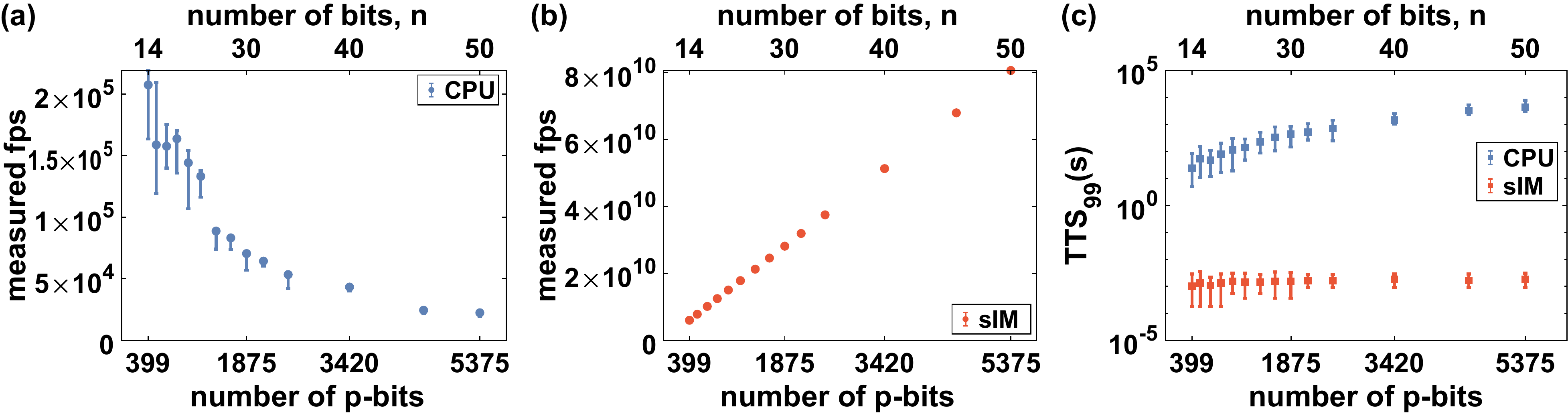}
    \caption{Performance scaling between parallel (sIM) and serial (CPU) implementations. (a) The CPU flips per second (fps) as a function of graph size is shown, larger graph sizes drastically limiting the fps. (b) The measured sIM  fps as a function graph size is shown, showing ideal parallelism scaling linearly with system size in contrast to the serial CPU. (c) Time to solution comparison for approximate factorization between the CPU and the sIM as a function of graph size. $\rm TTS_{99}$ or time to solution, is defined as the time to reach 99\% of the ground state.  $\rm TTS_{99}$ requires up to 4408.93 s for the 50-bit factorization in the CPU. For the sIM, it is showing almost a flatline behavior with up to $\approx 10^{6}$x performance improvement.} 
    \label{fig:CPU_sIM_FPS_TTS}
\end{figure*}

\subsection{CPU comparison with approximate factoring}
\label{sec:cpu_fact}

In order to test the parallelism achieved by our approach, first, we compare the sIM implementation of \emph{parallelized} Gibbs sampling with a CPU implementation of standard (serialized) Gibbs sampling \cite{koller2009probabilistic}. Our purpose in this comparison is to stress the asymptotic scaling differences between the standard approach and our method, rather than a pure performance comparison which we perform later in Section \ref{sec:Benchmarking}, against highly optimized and fine-tuned GPU/TPU implementations. 

We define approximate factorization as reaching 99\% of the absolute ground energy from 14-bit to 50-bit semiprimes. We do not attempt exact factorization since the absolute ground state is very difficult to be reached using simulated annealing \cite{aarts1989simulated} and the CPU practically never reaches there.  While finding approximate factors is not useful for the problem of integer factorization, many other optimization problems benefit from high-quality and approximate solutions. As such, we treat approximate factorization as a computationally hard problem benchmark.  

We use the same annealing schedule and the same sparsified graphs for the comparison between an unoptimized serial MATLAB implementation on a CPU and the parallel FPGA implementation of the sIM. For each problem, we have attempted to factor 10 different numbers 10 times to make sure we collect enough statistics and test the robustness of the system. 

As a key metric, we first focus on the flips per second (fps) as a figure of merit. The fps corresponds to the \emph{correlated} flips per second that can be taken by the system where every flip is made based on the latest state of the network, between physically connected nodes while avoiding simultaneous updates. Indeed, various hardware implementations for MCMC solvers have reported this metric \cite{block2010multi, preis2009gpu,yang2019high,fang2014parallel}. The fps can be thought of as the effective processing speed for MCMC. The key point of the parallel architecture we design for the sIM  is that its fps \emph{increases linearly} with the number of p-bits corresponding to a problem.

Fig.~\ref{fig:CPU_sIM_FPS_TTS}a,b present a comparison of fps between the inherently serial CPU  and the massively parallel sIM  for the factorization problem. The CPU calculation is done in MATLAB by an iterative solution of Eq.~\eqref{eq:syn}-\eqref{eq:pbit} using standard Gibbs sampling \cite{koller2009probabilistic} with simulated annealing. Even though optimization techniques including graph coloring by multiple threads could improve our MATLAB implementation, our purpose is to stress the massive parallelism achieved by the sparse architecture we develop over a standard implementation of Gibbs sampling. 

 MATLAB runs on the Knot Cluster at the Center for Scientific Computing (CSC) server, UCSB featuring an Intel Xeon Processor E5630 running at up to 2.8 GHz. On the other hand, we have used five equally phase-shifted 15 MHz clocks in the sIM  and assigned those clocks to the p-bits based on a previously calculated graph coloring. In the sIM, each p-bit updates in parallel, achieving an effective clock speed of $N \times$15 MHz where $N$ is the number of p-bits. 15 MHz is chosen to satisfy the timing requirements for the additions to be completed and better FPGAs or hardware implementations can be envisioned to reach even higher frequencies. 

It is important to note that we directly measured the flips per second of the sIM  by means of specially designed reference counters in the FPGA (See the Methods Section~\ref{methods_fps} for the measurement details). The maximum flips per second (fps) achieved by the CPU is limited to around $10^{5}$ (Fig.~\ref{fig:CPU_sIM_FPS_TTS}a). Moreover, fps quickly starts to decline for the CPU with the increasing number of p-bits. The main reason for this is due to the sequential nested for-loops in standard Gibbs sampling \cite{koller2009probabilistic}. On the contrary, the sIM  collects up to 80 - 100 billion fps for the highest problem sizes. Crucially, the fps increases linearly with the increasing number of p-bits as shown in Fig.~\ref{fig:CPU_sIM_FPS_TTS}b due to its massively parallel architecture. Starting at an fps of 5.99$\times 10^{9}$ for the 14-bit factorization, it achieves a maximum fps of 8.06$\times10^{10}$ for the 50-bit factorization. 

While we do measure an increasing fps for the sIM  as a function of graph size, an important arising question is whether all these samples are useful or not. In order to test this question, we define a performance metric, the time to solution, $\rm TTS_{99}$ as the time to reach 99\% of the absolute ground energy. Note that in the case of factorization, the exact solution is planted, in other words, we know the factors of a given product and therefore we have access to the exact ground state energy.
Fig.~\ref{fig:CPU_sIM_FPS_TTS}c shows a steep rise of $\rm TTS_{99}$ from 14-bit to 50-bit factorization for the CPU. The fastest case requires $\rm TTS_{99}$ of 23.84 s for the 14-bit factorization while the slowest one requires 4408.93 s for the 50-bit factorization. In contrast, the sIM shows a roughly constant mean value for all the problems, requiring 1.02 ms for the 14-bit factorization and  1.84 ms for the 50-bit factorization (See the Methods Section~\ref{methods_TTS} for TTS measurement details). The 50-bit factorization is over a 2.4$\times10^{6}$x improvement over the CPU. The difficulty of the approximate factorization is clearly increasing with increasing problem size. The reason for the constant time to solution for the sIM  despite the increasing difficulty of the problem for larger problem sizes can be attributed to the massive parallelism of the sIM  where increasing problem size also increases the fps. Therefore, we can conclude that the measured fps shown in Fig.~\ref{fig:CPU_sIM_FPS_TTS}b is a real improvement. This makes the massive parallelism of the sIM  very different from trivially parallel p-bits sampling independently.

\subsection{GPU and TPU comparison}
\label{sec:Benchmarking}
Table~\ref{tab:benchmarking} summarizes the performance benchmarking of the current work (sIM) with the state-of-the-art GPUs and TPUs. An important distinction between these comparisons is that virtually all optimized implementations of GPUs and TPUs make use of a regular 2D chessboard lattice where partitioning the graph into two color blocks becomes the key piece that enables parallelism. By contrast, in our examples, we show that for realistic instances of combinatorial optimization problems using invertible Boolean circuits, the resulting graphs are sparse but not necessarily regular or bipartite (2-color). Even though theoretical results suggest simple nearest-neighbor models could be sufficient to model any other problem \cite{de2016simple}, accelerating problem graphs in between nearest-neighbor and all-to-all will be critically important in practice. 

The performance of the sIM flips per second (fps) grows linearly with the number of p-bits in a network, hence the fps becomes a size dependent metric. We observe the sIM reaches up to 143.8 flips/ns at one of the largest graph nodes (4793 p-bits). This is an 18.03x performance gain over the single Nvidia Tesla C1060 GPU with multi-spin coding \cite{block2010multi, preis2009gpu}. It also outperforms the Google Cloud TPU implementation of the 2D Ising model by a factor of 11.17 and its reference Nvidia Tesla V100 GPU by a factor of 12.65 \cite{yang2019high}. Furthermore, the sIM provides 4.82x more flips/ns than the Nvidia Fermi GPU coded for simulating the 3D Edwards–Anderson model with parallel tempering \cite{fang2014parallel}.

\begin{table}[t!]
    \centering
    \begin{tabular}{|c|c|c|}
        \hline
        {\bf Platform} & {\bf Graph} & {\bf flips/ns} \\
        \hline 
        \hline 
        Nvidia Tesla C1060 GPU \cite{block2010multi, preis2009gpu} & Chessboard & 7.98 \\
      
        Nvidia Tesla V100 GPU \cite{yang2019high} & Chessboard & 11.37 \\
       
        Google TPU \cite{yang2019high} & Chessboard & 12.88 \\
       
        Nvidia Fermi GPU \cite{fang2014parallel} & Chessboard & 29.85 \\
        
        \hline
        \hline 
         FPGA sIM [This work] &  Irregular &  143.80\\
         Nanodevice sIM [Projected] \cite{sutton2020autonomous,borders2019integer,hayakawa2021nanosecond,safranski2021demonstration} &  Irregular &  1,000,000\\
        \hline
    \end{tabular}
    \caption{Comparison of the FPGA-based and nanodevice-based (projected) sIM with optimized GPU and TPU implementations of Markov Chain Monte Carlo sampling. Unlike the GPUs and TPUs, the sIM can support regular chessboard lattices as well as irregular graphs shown in this paper. For the sIM, fps is a size dependent metric as shown in Fig.~\ref{fig:CPU_sIM_FPS_TTS},\ref{fig:winner_solvers}. In this table, the best fps achieved in the largest system is quoted.}
    \label{tab:benchmarking}
\end{table}

Beyond the FPGA-based sIM implementation we consider in this paper, we make performance projections for massively parallel asynchronous sIMs using nanodevices (see Supplementary Section~\ref{sec:characteristic_devices}). Magnetic Tunnel Junctions (MTJ) have recently attracted attention as building blocks for probabilistic computation because of their extreme scalability. The magnetic memory industry have integrated up to billions of single MTJs to replace various parts of the memory hierarchy \cite{bhatti2017spintronics}. Through minimal modifications  such MTJs can be made stochastic \cite{borders2019integer,kobayashi2021sigmoidal}, providing the expensive random number generation with negligible hardware cost. Stochastic MTJs have been demonstrated to provide fast fluctuations ($\tau=$1 ns / flip) \cite{safranski2021demonstration,hayakawa2021nanosecond} and at least a million MTJs ($N=10^6$) can be integrated in massively parallel architectures \cite{sutton2020autonomous,borders2019integer} similar to what we consider in this paper. Following the linear scaling law we demonstrated in Fig.~\ref{fig:CPU_sIM_FPS_TTS}b, such sIMs can provide $N/\tau$ flips per second reaching 1 million flips per nanosecond (Table~\ref{tab:benchmarking}), provided that the connectivity of the hardware is sparse enough to enable the ideal parallelism demonstrated in this paper. 

\begin{figure*}[!t]
    \centering
    \includegraphics[width=0.735 \textwidth]{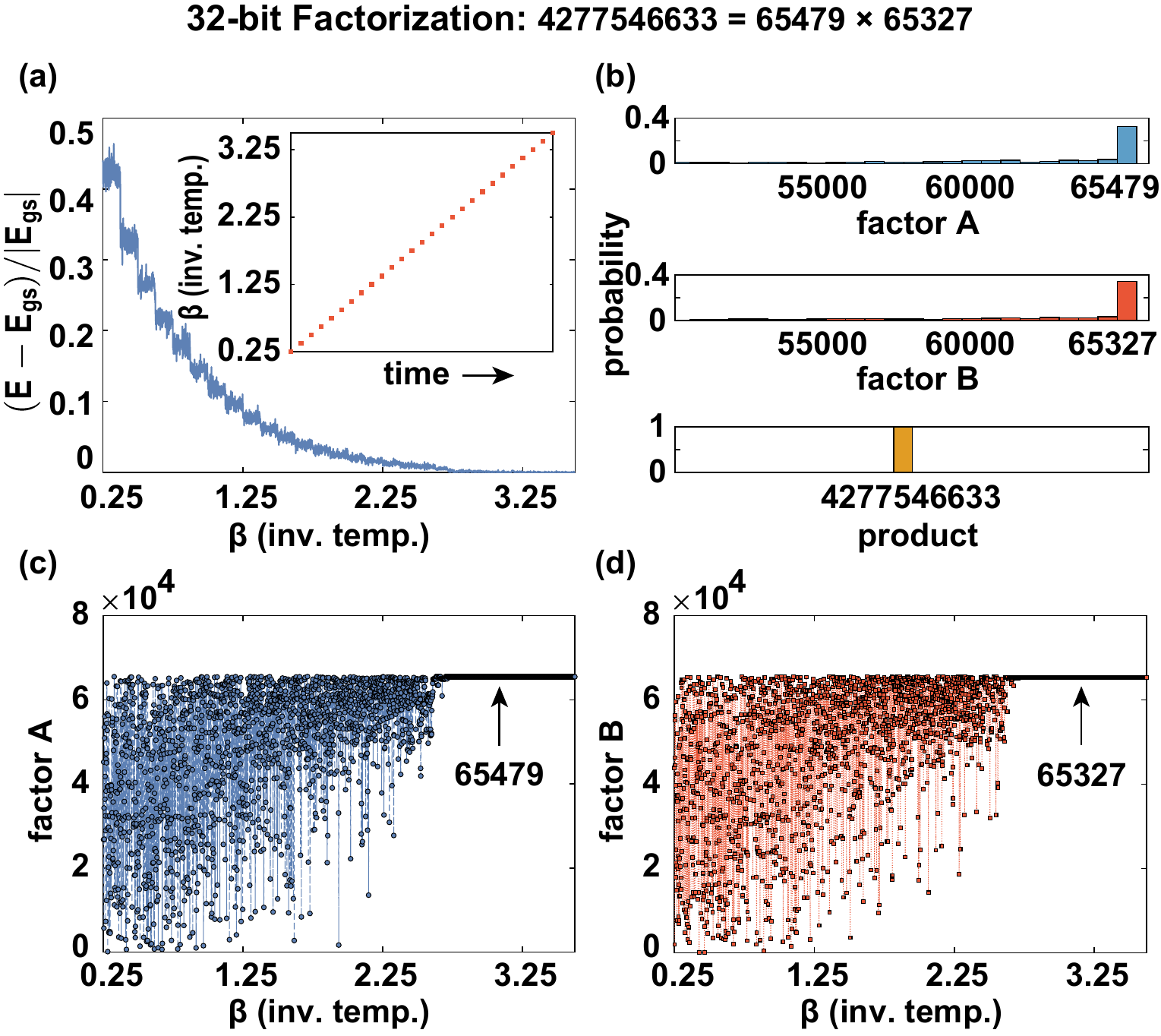}
    \vspace{10pt} 
    \caption{Exact factorization of a 32-bit number, P= 4277546633 in the sIM. (a) The normalized energy  as a function of inverse temperature, ($\beta$ from Eq.~\eqref{eq:boltz}. Inset shows the linear annealing schedule. (b) Histograms of the product P and the factors A, B over the entire annealing schedule. (c-d) Factors A and B at different $\beta$ values, showing they converge to the right ground state at the highest $\beta$.}
    \label{fig:32-bit}
\end{figure*}

\section{Exact Factorization of Semiprimes up to 32-bits}
\label{sec:exact_fact}

\begin{table}[t!]
    \centering
    \begin{tabular}{|c|c|}
        \hline
        {\bf Platform} & {\bf Integers Factored (up to)}  \\
        \hline 
        \hline 
        D-Wave 2000Q \cite{andriyash2016boosting} & 143 (8-bit)\\
        CMOS Inv. Logic \cite{smithson2019efficient} & 598 (10-bit) \\
        Stochastic MTJ \cite{borders2019integer} & 945 (10-bit)\\
        FPGA RBM \cite{patel2020logically} & 43621 (16-bit)\\
        D-Wave 2000Q \cite{dridi2017prime} & 223357 (18-bit)\\
        D-Wave 2000Q \cite{jiang2018quantum} & 249919 (18-bit)\\
        \hline
        \hline 
         FPGA sIM [This work] &  4277546633 (32-bit) \\
        \hline
    \end{tabular}
    \caption{Comparison between the state-of-the-art hardware factorizers and the FPGA-based sIM. Note that all of these solvers treat integer factorization as a frustrated spin-glass problem and perform classical or quantum annealing (or ordinary sampling). We show best reported numbers and single instances, in the case of sIM, random semiprimes up to 32-bits can be reliably factored, see Section~\ref{sec:exact_fact}.}
       \label{tab:factors}
\end{table}

Beyond approximate factorization, we have also performed exact factorization with the sIM. As mention in Section~\ref{sec:summary_results}, we consider the integer factorization problem as a benchmark to compare the performance of our sIM implementation against other probabilistic solvers or D-Wave's quantum annealers.  We found that the sIM can factor random semiprimes up to 32-bits reliably (Supplementary Information, Fig.~\ref{fig:TTS_100}). To the best of our knowledge, this result is by far the best among all other approaches of solving factorization as an optimization problem, for example by D-Wave and others \cite{andriyash2016boosting, borders2019integer, smithson2019efficient,dridi2017prime, jiang2018quantum, patel2020logically}. Table~\ref{tab:factors} presents a comparison of integer factorization across the state-of-the-art hardware platforms, and the sIM reports the largest factorization up to 32-bit. What is common to all these solvers is they express factorization as a frustrated spin-glass problem in which the ground state is searched using classical or quantum annealing. 

From an algorithm perspective, factorization of 32-bit semiprimes is not difficult since even with trial division this is a relatively easy computation. From a statistical physics perspective, however, finding the doubly degenerate ground state of a frustrated spin glass in a $2^N$ dimensional space ($N>2000$ p-bits) is striking. Contrasting the time to solution shown for approximate factorization (Fig.~\ref{fig:CPU_sIM_FPS_TTS}c) to that of exact factorization (Fig.~\ref{fig:TTS_100}) as  function of problem size, we observe a drastic difference in algorithmic scaling, indicative of a ``golf course'' like energy landscape where the ground state is well-hidden from ``ordinary'' approximate states that are easy to reach. 

It is worth stressing that we  used a simple, standard simulated annealing algorithm without any fine tuning or optimization. Our preliminary findings  indicate parallel tempering or other algorithmic methods could improve the success probability of these results. We believe the success of our approach over similar alternatives is due to the sparsification methods that have enabled the massively parallel sampling architecture of the sIM.

Fig.~\ref{fig:32-bit} presents the exact factorization of a 32-bit number, P= 4277546633. The linear annealing schedule and the normalized energy of the system are presented in Fig.~\ref{fig:32-bit}a. The absolute ground state is reached at the coolest temperature (the highest $\beta$). The histograms of the product P and the factors A, B over the entire annealing schedule are presented in Fig.~\ref{fig:32-bit}b where the exact factors A = 65479 and B = 65327 are visited reliably. Fig.~\ref{fig:32-bit}c and Fig.~\ref{fig:32-bit}d reconfirm that the factors are consistently found at the highest $\beta$ without any fluctuations. 

While we do not show statistics in Fig.~\ref{fig:32-bit}, in Supplementary Fig.~\ref{fig:TTS_100}, we report the time to find the exact factors ($\rm TTS_{100}$) from 14-bit to 32-bit semiprimes. For any of these problems, the CPU fails to find the exact factors even over a very long time and therefore is excluded from the $\rm TTS_{100}$ report and we report the sIM times. As before, we attempt to factor 10 different numbers 10 times for each problem. 

Unlike approximate factorization where we defined $\rm TTS_{99}$ as the average time the sIM takes before reaching 99\% of the absolute ground state, we find that for exact factoring there is an (empirical) exponential dependence of the time with respect to problem size, in line with the belief that integer factorization is in NP where a known polynomial algorithm does not exist \cite{lenstra1990number}. 

To reiterate, extrapolating our observed data with an exponential fit to estimate exact factoring, we find that factorization with this method is not a practical approach in the context of cryptography, in agreement with the observation from Ref.~\cite{mosca2019factoring}. However, improving SAT solving with massively parallel hardware could still be useful in accelerating critical subroutines of the best factoring algorithms \cite{mosca2020speeding}.  

\begin{figure*}[t!]
    \centering
    \includegraphics[width =0.875 \textwidth ]{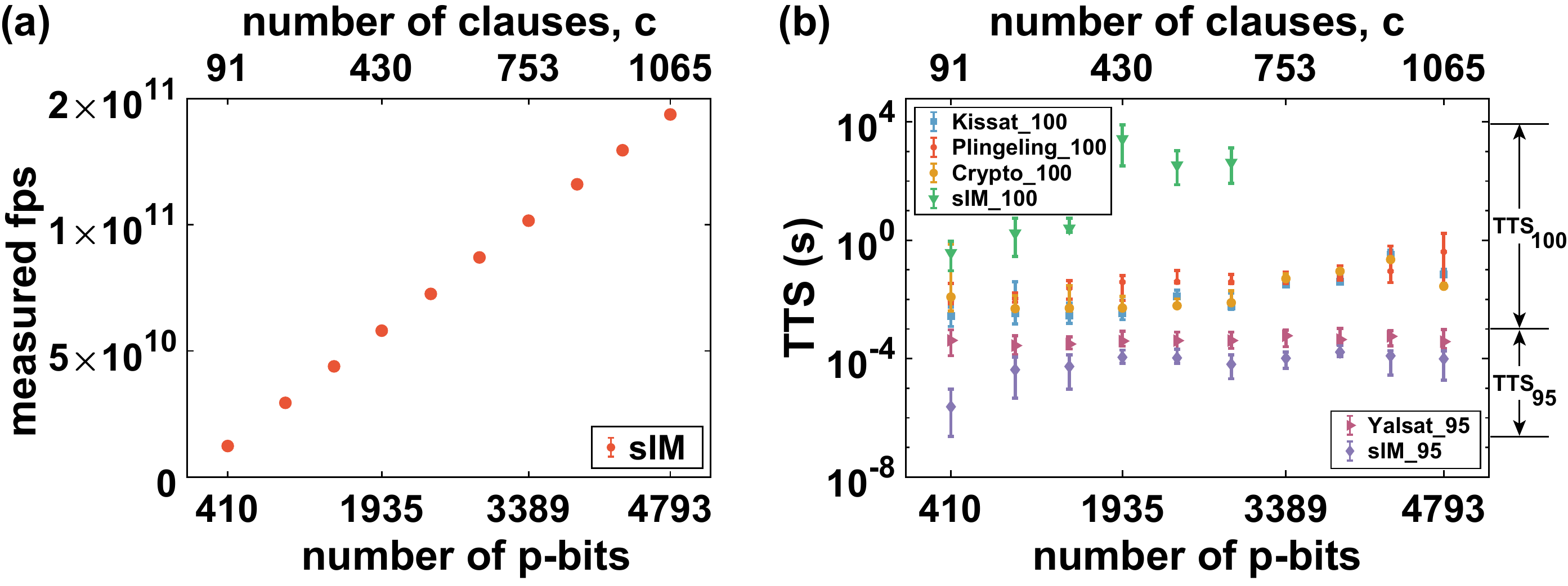}
    \vspace{10pt} 
    \caption{Performance of the sIM vs. competition-winning SAT solvers. (a) The flips per second (fps) of the sIM increases linearly with the graph size, showing ideal parallelism. The sIM  achieves a record fps of 1.44$\times10^{11}$ for the largest problem with 4793 p-bits and 1065 clauses. (b) Runtimes to solve the UBC SATLIB \cite{satlib} 3SAT instances. Winning solvers from the SAT 2020 competition, namely Kissat,  Plingeling,  and,  Cryptominisat \cite{kissat, cryptominisat} find 100\% solution (all clauses satisfied). SAT competition 2017 random track winner Yalsat \cite{Yalsat} is set to find the 95\% solution (95\% of the clauses satisfied). In exact SAT solving, the sIM is slower than all professional SAT solvers. However, in approximate SAT solving to satisfy 95\% of the clauses, the sIM outperforms all of these solvers (including local search-based SAT solvers, such as Yalsat) delivering the fastest solution.}
    \label{fig:winner_solvers}
\end{figure*}

\section{Boolean Satisfiability with Invertible Logic}
\label{sec:SAT}

\vspace{-2mm}
\subsection{sIM vs. competition-winning SAT Solvers}

As shown earlier in Fig.~\ref{fig:fig1}, it is possible to design invertible Boolean circuits in hardware corresponding to satisfiability instances using the principles of invertible logic. Here, we focus on solving 3SAT problems to demonstrate the hardware acceleration of our massively parallel sIM architecture. In particular, our purpose is to compare the sIM with the best possible software algorithms and we focus on competition winning SAT solvers as a benchmark. 
As previously, we first report the flips per second (fps) achieved by the sIM for different 3SAT instances
defined by the number of their clauses (Fig.~\ref{fig:winner_solvers}a). The sIM runs with four parallel and equally phase-shifted clocks operating at 30 MHz since in this case the sparsified graphs require only 4 colors. For the smallest instance `uf20-01.cnf' with 20 variables and 91 clauses, the sIM achieves an fps of 1.23$\times10^{10}$ with 410 p-bits. For the largest instance `uf250-01.cnf' with 250 variables and 1065 clauses, the sIM achieves a record fps of 1.44$\times10^{11}$ with 4793 p-bit, a 5 to 18x speed up over optimized TPUs and GPUs  discussed in Section~\ref{sec: Results_fact} (Table~\ref{tab:benchmarking}).

Fig.~\ref{fig:winner_solvers}b shows the run times to solve the UBC SATLIB \cite{satlib} 3SAT instances using different professional SAT solvers and the sIM. We solve each instance 100 times to obtain enough statistics. We compare our results with award-winning solvers from the SAT 2020 competition, namely Kissat, Plingeling, and Cryptominisat \cite{kissat, cryptominisat}. These conflict-driven clause learning (CDCL) solvers attempt to find the exact solution that satisfies all of the clauses. All these solvers are executed on the same Linux machine having a flagship Intel Core i9-10900 Processor running at up to 5.20 GHz. Time to solve all the clauses (100\% solution) is reported as $\rm Kissat_{100}$, $\rm Plingeling_{100}$, and, $\rm Crypto_{100}$ respectively. We have used linear simulated annealing in the sIM to report the time to solve all the clauses, labeled as $\rm sIM_{100}$ in Fig.~\ref{fig:winner_solvers}b.

As typical of simulated annealing \cite{aarts1989simulated}, reaching the absolute ground state is difficult for the sIM. We do report the $\rm sIM_{100}$, namely the time it takes for sIM to satisfy all clauses in a given 3SAT instance and find that despite the enormous number of flips per second taken by the massively parallel processor, we did not find the ground state beyond 2903 p-bits (Fig.~\ref{fig:winner_solvers}b, $\rm sIM_{100}$).   

In many practical instances however, the user may not be interested in finding the absolute ground state of an optimization problem, and reaching approximate but practically useful solutions as quickly as possible is far more important. In such a paradigm, we find that the sIM beats all of the SAT solvers mentioned (Fig.~\ref{fig:winner_solvers}b, $\rm sIM_{95}$). Because CDCL-based solvers such as Kissat, Plingeling and Crpytominisat are programmed to find the exact solution, we also test the sIM against another solver Yalsat (2017 SAT competition random track winner) which keeps a current best solution around. We program the solver to stop when it reaches 95\% of the solution (denoted as $\rm Yalsat_{95}$). We run Yalsat on the same Linux Machine. sIM is also set to solve 95\% of all the clauses and the time to solution is noted as $\rm sIM_{95}$. We find that in this approximate SAT solving mode, the sIM provides the fastest approximate solution outperforming all of the professional SAT solvers by a factor of 4 to 700. We find that there is no failure even for the largest instance (`uf250-01.cnf') we can fit to our FPGA encoded with 4793 p-bits. It takes only 2.36 $\mu$s for the instance `uf20-01.cnf' and 98.26 $\mu$s for the instance `uf250-01.cnf' to solve 95\% of the clauses. We expect larger improvements in more scaled implementations of our sIM architecture using more powerful FPGAs or application specific integrated circuits.   

\begin{figure*}[t!]
    \centering
    \includegraphics[width=0.75 \textwidth]{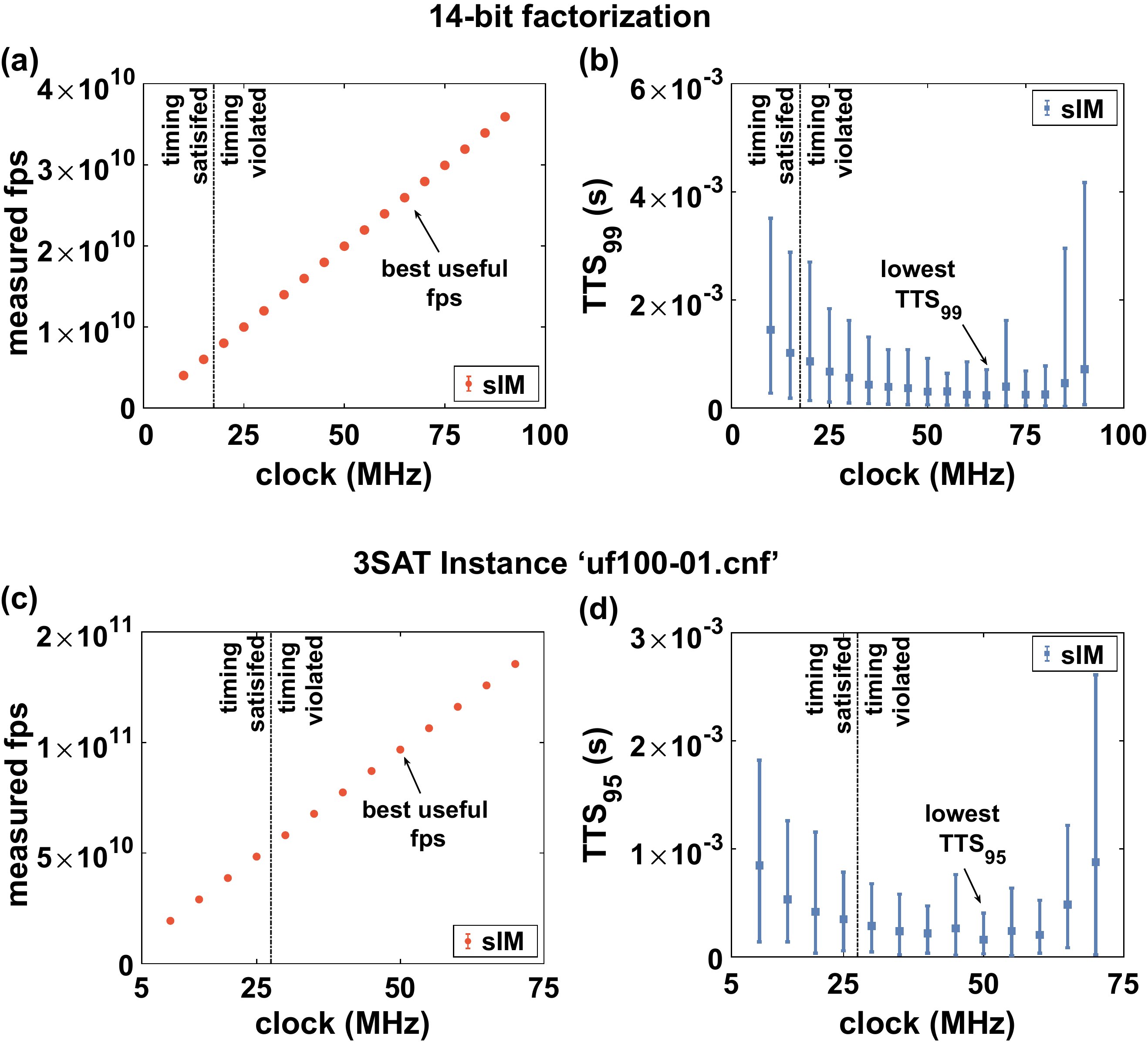}
    \caption{Overclocked Gibbs Sampling with sIM. (a)  14-bit factorization: the flips per second (fps) as a function of p-bit clocks is shown. Timing is violated after 15 MHz. Based on (b), we observe that  overclocking effectively improves fps up to 65 MHz. (b) Time to solution to reach 99\% of the ground state, $\rm TTS_{99}$, measured as a function of clock frequency. (c) 3SAT instance `uf100-01.cnf': measured fps as a function of p-bit clocks shown. Timing is violated above 25 MHz. Based on the results of (d), the highest effective fps is achieved at 50 MHz. (d) Time to solution to satisfy 95\% of the clauses, $\rm TTS_{95}$ as a function of clock frequencies. Both results (b)-(d) show qualitatively similar behavior of improving performance followed by a sharp decline, indicative of universal behavior.} 
    \label{fig:overclock}
\end{figure*}
\section{Overclocked Gibbs Sampling with Sparse Ising Machines}
\label{sec: overclocking}
The parallelized sparse Ising Machine architecture we introduced in this work implements Eq.~\eqref{eq:syn}-\eqref{eq:pbit} exactly. This is ensured by making every color block update with the most up-to-date neighbor information. Since this architecture is inspired by asynchronous and physical p-bit implementations, for example using stochastic magnetic tunnel junctions, a natural question to consider is whether inexact Gibbs sampling where the p-bits do not update with the most up-to-date neighbor information is worth considering. This approach is reminiscent of asynchronous Gibbs sampling (or Hogwild! Gibbs) approaches that have been analyzed theoretically \cite{johnson2013analyzing,daskalakis2018hogwild,de2016ensuring}. 

Here, we (systematically) investigate how increasing individual clock frequencies of color blocks and eventually performing \emph{inexact} Gibbs sampling with old statistics at the p-bit level affects system performance. Remarkably, for two completely different problems (integer factorization and Boolean satisfiability), we observe qualitatively similar results as a function of increasing clock frequency, therefore increasing the number of messages dropped between neighbors, reminiscent of approximate message passing algorithms \cite{mezard2009information}. In both cases (Fig.~\ref{fig:overclock}b,d), we observe an initial decrease in time to solution with increasingly incorrect updates followed by a sharp increase. Increasing the clock frequencies naturally increase the fps of the main network (Fig.~\ref{fig:overclock}a,c), however, when messages are dropped beyond a certain threshold, the improvement in the fps does not help the network converge to the right answers. 

To test the generality of overclocking, we analyzed inexact Gibbs sampling with systematically introduced errors in a 5 p-bit full adder circuit  (Supplementary Section~\ref{sec:mask_model}).  By introducing two different error models, we  observe qualitatively similar behavior, where introducing a small amount of error (related to the number of messages dropped between neighbors) does not lead to significant deviations from the exact Boltzmann distribution. This explains why a moderate amount of overclocking is effective: increasing fps without introducing significant errors decreases time to solution, as observed in two different problems in Fig.~\ref{fig:overclock}. In Supplementary Section~\ref{sec:mask_model}, we also show how further overclocking reduces the network to a fully synchronous (parallel) updating state, providing analytical estimates of limiting behavior.  

One additional reason why overclocking improves performance significantly is due to the slowing down of the network dynamics at lower temperatures (higher $\beta$). At the end of an annealing schedule, despite 
introducing timing failures in many critical paths (and potentially dropping a significant number of messages between neighbors), these critical paths are not activated because p-bits do not change their states frequently. The degree of resilience of the system to errors through inexact sampling is an important feature of such probabilistic methods \cite{zhang2021statistical} which can be exploited in truly asynchronous nanodevice-based implementations of sIM \cite{sutton2020autonomous,borders2019integer,hayakawa2021nanosecond,safranski2021demonstration}. 

\section{Conclusions}
\label{sec:conclusions}
In this paper, we proposed and implemented a massively parallel architecture, the sparse Ising Machine, to parallelize a broad range of Markov Chain Monte Carlo (MCMC) algorithms useful for computationally hard problems. In particular, overcoming the fundamentally serial nature of MCMC algorithms such as Gibbs sampling, we have shown an architecture that can achieve ideal parallelism where the main metric of the sIM, the flips per second, scales linearly with the number of probabilistic bits in the system. This parallel architecture used several algorithmic ideas of combining invertible logic to produce sparse graph representations of combinatorial optimization problems such as Boolean satisfiability and integer factorization. Further sparsification was needed to ensure matrix multiplication and addition can be performed before independent p-bits are updated. The architecture used approximate graph coloring to parallelize sampling. 

We have shown an FPGA-based implementation of this concept where we have achieved three major results. First, comparisons to an ordinary CPU implementation of Gibbs sampling showed that the sIM is able to achieve up to 6-orders improvement in flips per second, which directly translated to advantages in time to solution in the integer factorization problem. Comparisons to highly optimized GPU and TPU implementations, the sIM showed up to 5-18x measured speed up in flips per second, without the use of regular or simple graphs as commonly used for benchmarking purposes in GPUs and CPUs. Second, the sIM was able to factor semiprimes up to 32-bit integers, far larger than the best available results on factoring where an optimization approach is taken. And third, the sIM was able to beat competition winning SAT solvers in approximate satisfiability, delivering superior performance compared to the best possible classical approach in solving satisfiability problems. We have also shown how overclocking in the spirit of asynchronous Gibbs sampling \cite{johnson2013analyzing} could lead to performance improvements. 

These results were obtained in a FPGA platform where our problem sizes were limited to the number of probabilistic bits we could fit in a single device. Use of more powerful FPGAs would immediately extend the size of problems programmable to the sIM. The ideal parallelism we achieved in the architecture, coupled with algorithmic sparsification techniques we developed can further be exploited in highly scaled implementations. In particular, nanodevice (or analog CMOS-based) p-bits can produce significant improvements over our present results, as the search for domain-specific hardware in the beyond Moore era of electronics intensifies. 

\section{Methods}
\label{methods}

\subsection{Problem description}
\label{sec:prob_descript}
For the factorization problem, we generated random semiprime numbers from 14-bit to 50-bit using MATLAB. For each instance, 10 different numbers were generated. The graphs obtained using invertible logic and sparsification are very sparse (See Supplementary Fig.~\ref{fig:density}a). 

For the SAT problem, we solved 3SAT instances (each clause has exactly 3 variables). The instances were  collected as .CNF files from the UBC SATLIB library \cite{satlib}. Similar to factorization, the 3SAT graphs are very sparse (See Supplementary Fig.~\ref{fig:density}b).

\subsection{Simulated annealing}
In simulated annealing,  $\beta$ is gradually increased over time. According to Eq.~\eqref{eq:syn}, multiplication of $\beta$ and the input weights ($J$, $h$) are performed in MATLAB. The updated values of $J$ and $h$ are sent to the FPGA for every $\beta$ over time to do simulated annealing.

\subsection{Data READ/WRITE}
MATLAB is used to READ/WRITE data from the FPGA through a USB-JTAG interface (see Supplementary Fig.~\ref{fig:architecture}a). A programmable timer is implemented in the FPGA. Using the timer, a global DISABLE signal is sent to the p-bits before a READ instruction. The timer is preset from the program (MATLAB) and all the p-bits are automatically frozen at the same time when the time is up. Once the p-bits are frozen, the data are READ using the USB-JTAG interface and sent to MATLAB for post-processing. When the READ instruction is DONE, the timer is RESET from MATLAB to resume the p-bits if necessary. Similarly, for the WRITE instruction, a global DISABLE signal is sent using the programmable timer to freeze the p-bits before sending the weights. Likewise the READ instruction, the timer is RESET from MATLAB to resume the p-bits after the WRITE instruction is DONE. 

\subsection{Measurement of fps}
\label{methods_fps}
Each p-bit is designed with a programmable stopwatch counter in the FPGA. A global counter running parallelly is set to count up to a preset value at the positive edge of a known clock. When the global counter is DONE counting, a global DISABLE signal is broadcast to all other counters. Comparing the p-bit counter outputs (number of flips) with the global counter preset value, the time for the total flips is obtained. With this data, the fps of the sIM is measured experimentally for each p-bit. To measure the fps in the case of the CPU, built-in functions from MATLAB is used to measure the elapsed time and programmatically count the total flips in that time. With this data, the fps is measured in real-time. The error bars in all the figures are obtained by taking 100 measurements of fps.

\subsection{Measurement of TTS}
\label{methods_TTS}
In the FPGA, a minimum time is set using the programmable timer to find the solution to the problem of interest. After that time, a global DISABLE signal is sent to READ the latest TTS. In iterations, the minimum time is incremented, and the p-bits are RESET. This process is repeated until the desired solution is reached. The latest TTS is reported as the TTS of the sIM for that problem. In measuring the TTS, we do not include the READ/WRITE times through the USB-JTAG interface. While we use  a slow USB-JTAG interface (up to 33 MHz) for the convenience of using MATLAB, much faster R/W protocols such as PCI Express (up to 8 Gb/s) would remove this time entirely.  To measure the TTS in the case of the CPU, a predefined minimum number of samples is set to find the solution. The number of samples is increased in iterations until the optimum solution is found by the CPU. The time to solution is recorded using the built-in function and the latest one is reported as TTS of the CPU for that problem. The error bars in all the figures are obtained by taking 100 measurements of TTS.

\subsection{Setting up the SAT solvers}
The online source codes of the SAT solvers are used to build the solvers on a Linux machine. For the CDCL SAT solvers, the time to find the 100\% solution is measured using a simple Python script. For the local SAT solver Yalsat, the program is set to report the TTS for the current best solution. 

\begin{acknowledgments}
 The authors are grateful to Brian M. Sutton, Daniel Eppens, Alan Ho and Masoud Mohseni for useful discussions. It is a pleasure to acknowledge Xilinx for hardware support. K.Y.C. and L.T. acknowledge support from the Institute of Energy Efficiency, UC Santa Barbara.  K.Y.C. and N.A.A. acknowledge National Science Foundation support through CCF 2106260. The research of A.G., M.C., and G.F. has been supported by the Project No. PRIN 2020LWPKH7 funded by the Italian Ministry of University and Research and by Petaspin association (\url{www.petaspin.com}). Use was made of computational facilities purchased with funds from the National Science Foundation (CNS-1725797) and administered by the Center for Scientific Computing (CSC). The CSC is supported by the California NanoSystems Institute and the Materials Research Science and Engineering Center (MRSEC; NSF DMR 1720256) at UC Santa Barbara.
\end{acknowledgments}

\section*{Competing interests}
J.M.M. is affiliated with Zyphra Technologies Inc., San Francisco, CA, USA. All other authors have no competing interests.

\section*{Data availability}
The data that support the plots within this paper and other findings of this study are available from the corresponding author upon reasonable request.

\section*{Code availability}
The computer code used in this study is available from the corresponding author upon reasonable request.

\balance

\clearpage
\onecolumngrid
\section*{Supplementary Information}
\beginsupplement

\subsection{Characteristics of CMOS and nanodevice based p-bits}
\label{sec:characteristic_devices}
\begin{figure*}[h]
    \centering
    \includegraphics[width =1 \textwidth ]{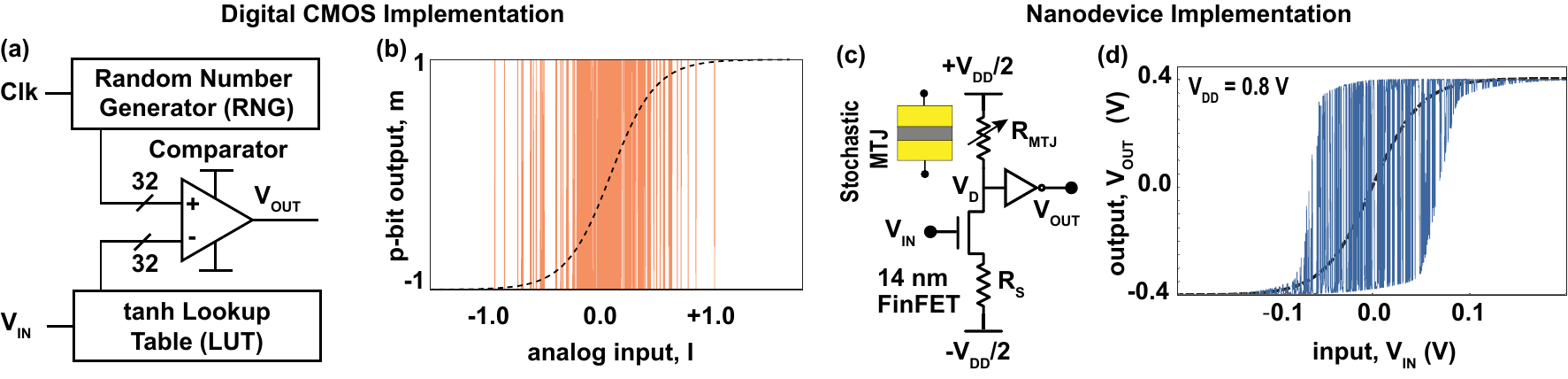}
    \caption{(a) Digital CMOS implementation of Eq.~\eqref{eq:pbit} of a p-bit using an RNG, a LUT, and a comparator. (b) Input-output characteristics of the digital CMOS p-bit. (c) Nanodevice implementation of Eq.~\eqref{eq:pbit} using a 14 nm FinFET and a stochastic MTJ. (d) Input-output characteristics of the nanodevice p-bit.}
    \label{fig:device}
\end{figure*}

In Section~\ref{sec:Intro} of the main paper, we have discussed Eq.~\eqref{eq:pbit} of a p-bit that can be implemented in several hardware platforms. A CMOS implementation of Eq.~\eqref{eq:pbit} is presented in Fig.~\ref{fig:device}a. A clock-triggered random number generator (RNG) provides the $\rm {rand}_U(-1,1)$ function and a lookup table (LUT) maps the tanh activation function. Finally, a comparator is used to trigger the flip of the p-bit. The input-output characteristics of a CMOS implemented p-bit is shown in Fig.~\ref{fig:device}b. Eq.~\eqref{eq:pbit} can also be implemented using nanodevices such as a 14-nm FinFET and a stochastic MTJ (see Fig.~\ref{fig:device}c) where a mapping between the dimensionless Eq.~\eqref{eq:pbit} and the device characteristics can be made \cite{faria2018implementing}. The input-output characteristics of the nanodevice p-bit is presented in Fig.~\ref{fig:device}d. The parameters used for the simulation are included in Table~\ref{tab:MTJ_parameters}. The simulation time steps to solve the stochastic Landau-Lifshitz-Gilbert equation intrinsically calculates an attempt time. For low-barrier nanomagnets, due to the lack of an energy-barrier there is no clear switching voltage at zero temperature, however, there is the notion of a `pinning current' \cite{hassan2021quant} which is a function of device parameters.

\begin{table}[h]
    \centering
    \begin{tabular}{|c|c|}
        \hline
        {\bf Parameter} & {\bf Value} \\
        \hline 
        \hline 
        Free layer energy barrier  &$\approx$ 0 kT \\
        Free layer diameter & 20 nm \\
        Free layer thickness   & 2 nm \\
        Free layer damping coeff. ($\alpha$) &  0.01 \\  
        $H_K$ (uniaxial anisotropy)& $10^{-2}$ Oe \\ 
        $H_D$ (demagnetization field)  & $4 \pi M_s$ \\
       Saturation magnetization ($M_s$) & 1100 emu/cc \\
        Interface Polarization &  $P=0.7$ \\ 
        Tunneling Magnetoresistance (TMR)  & $\displaystyle{2P^2}/{1-P^2}= 192\% $  \\ 
        Average conductance for MTJ ($G_0$) & 43 $\mu S$ \\ 
        NMOS technology model  & 14-nm HP FinFET, PTM \cite{zhao2007predictive} \\ 
         Time step for integration & $\Delta t=1$ ps\\
        \hline
    \end{tabular}
    \caption{Parameters used for nanodevice based p-bit simulation in Fig. \ref{fig:device}d.}
    \label{tab:MTJ_parameters}
\end{table}

\subsection{FPGA implementation of the sIM}
\label{sec:fpga_implement}
We have presented the experimental results of the sIM in the main paper without technical details about the implementation of the architecture. Here, we discuss an FPGA based implementation of the sIM in a Xilinx Virtex UltraScale+ VCU118 Evaluation board. The basic architecture of the FPGA design is presented in Fig.~\ref{fig:architecture}. 

\subsubsection{Interfacing unit}

We use MATLAB as an Advanced eXtensible Interface (AXI) master to communicate with the slave FPGA board through a USB-JTAG interface (Fig.~\ref{fig:architecture}a). We have designed an AXI master integrated IP on the board that transfers data with a 32-bit memory-mapped slave register IP via the fourth generation AXI (AXI4) protocol. An external website `airhdl' \cite{airHDL} is used to manage the memory mapping of the registers. 

\subsubsection{Clocking unit}

\begin{figure*}[t!]
    \centering
    \includegraphics[width =1 \textwidth ]{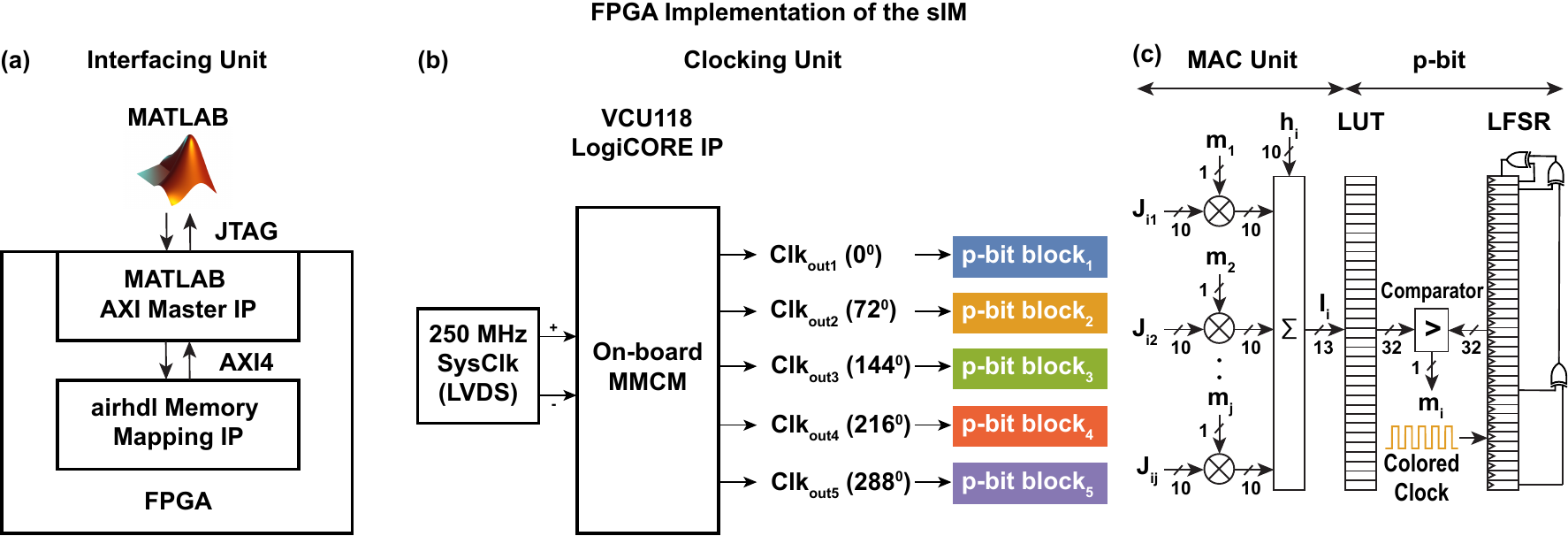}
    \caption{FPGA based implementation of the sIM. (a) An interfacing unit to communicate between MATLAB and the FPGA. (b) A built-in clocking unit to generate equally phase-shifted parallel clocks to trigger the colored p-bit blocks. (c) The MAC unit to implement Eq.~\eqref{eq:syn} where the colored clock from the general architecture is fed into the LFSR of the p-bit.} 
    \label{fig:architecture}
\end{figure*}

Section~\ref{sec: architecture_graph_coloring} in the main paper explains how approximate graph coloring can be used to color p-bit blocks for massive parallelism. For this, we have used built-in clocks on the FPGA board to drive the LFSRs inside the p-bit blocks as shown in Fig.~\ref{fig:architecture}b. A 250 MHz Low-voltage Differential Signaling (LVDS) system clock generates equally phase-shifted and parallel stable clocks using the on-board Mixed-Mode Clock Manager (MMCM) Module. This module is available in the VCU118 LogiCORE IP provided by Xilinx. The generated clocks are very accurate, have minimum jitter and minor phase error. The colored p-bit blocks get triggered with these phase-shifted clocks.

\subsubsection{MAC unit}
\label{sec:MAC}
In Section~\ref{sec:summary_results} of the main paper, we described the MAC unit interconnecting the p-bits and computing  Eq.~\eqref{eq:syn}. Fig.~\ref{fig:architecture}c illustrates the MAC unit implemented in the FPGA. In this work, we have used 32-bit  linear-feedback shift registers (LFSRs) with taps [32, 22, 2, 1] as RNGs after extensive experiments with different types of RNGs (LFSR, Xoshiro128+ \cite{blackman2018scrambled}, and the Mersenne Twister \cite{matsumoto1998mersenne}) with different bit-widths. The LUT bit-width is configured accordingly and a comparator compares the outputs of the LUT and the LFSR. The input weights ($J, h$) are programmable and we multiply them by $\beta$ from the MATLAB level to implement simulated annealing.

\subsubsection{Correspondence between binary and bipolar variables}
\label{sec:binary_bipolar}
Eq.~\eqref{eq:en}-\eqref{eq:pbit} presented in Section~\ref{sec:Intro} of the main paper use bipolar variables. It is more convenient to use binary variables for the FPGA-based implementation of the sIM. In the MAC unit, all the variables are calculated using binary notations where the final output for a p-bit is $m_i \in \{0, 1\}$. The bipolar to binary conversion is done using the following equations:  
\begin{equation}
\mathrm{J}_{\text {binary }}=2 \mathrm{J}_{\text {bipolar }}
\end{equation}
\begin{equation}
\mathrm{h}_{\text {binary }}=\mathrm{h}_{\text {bipolar }}-\mathrm{J}_{\text {bipolar }} \mathbb{A}
\end{equation}
where, $\mathbb{A}$ is an $[N\times1]$ vector of ones, and $N$ is the number of p-bits.
The activation function values stored in the LUT is converted from a bipolar representation to a binary representation by mapping $\mathrm {tanh}$ to $\mathrm {(1 + tanh)/2}$.

\begin{figure*}[t!]
    \centering
    \includegraphics[width =0.98\textwidth ]{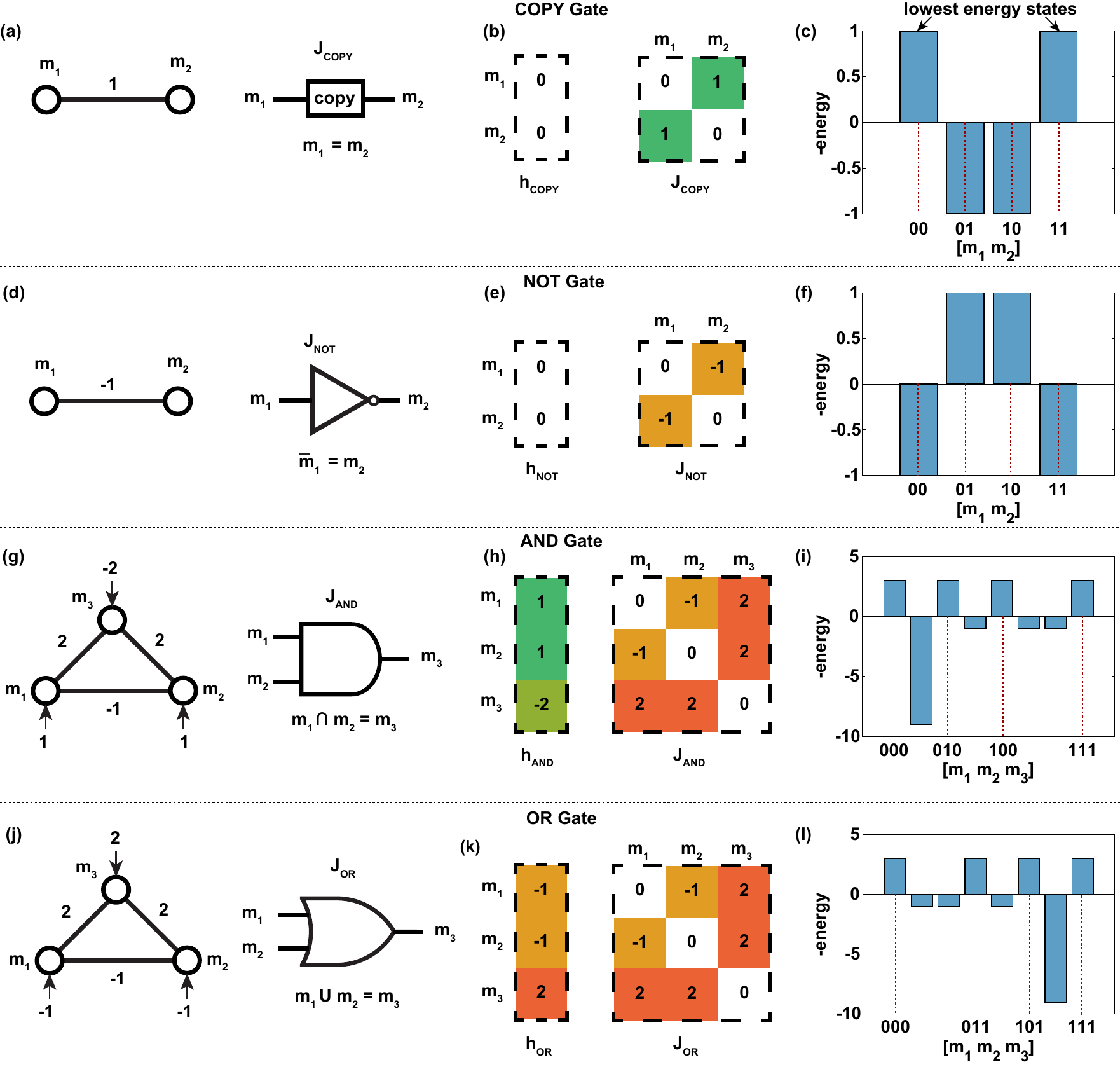}
    \caption{Basic logic gates for probabilistic computing. (a)-(l) COPY/NOT/AND/OR gates with $J$ and $h$ matrices having few unique weights highlighted using unique colors (not to be confused with graph coloring). The ground states match the basic truth tables where the vectors [$m_1~m_2$] or [$m_1~m_2~m_3$] are represented by the corresponding binary numbers.}
    \vspace{-2mm}
    \label{fig:gates}
\end{figure*}

\subsection{Basic logic gates for probabilistic computing}
\label{sec:basic}
Section~\ref{sec:composition} of the main paper illustrates how any invertible logic probabilistic circuit can be composed using basic logic gates and full adders to solve combinatorial optimization problems. Fig.~\ref{fig:gates}a-l presents the basic logic gates (COPY/NOT/AND/OR) used to build such p-circuits. The $J$ and $h$ matrices have few unique weights that are highlighted using unique colors. The energy plot corresponding to the Boltzmann probability are also included in the figure. For example, for the AND gate, the ground states (states with the lowest energy) $[m_1~m_2~m_3]$ = $\{000, 010, 100, 111\}_{\rm bin}$ correspond to the truth table of the AND gate.  

\begin{figure*}[t!]
    \centering
    \includegraphics[width =1 \textwidth ]{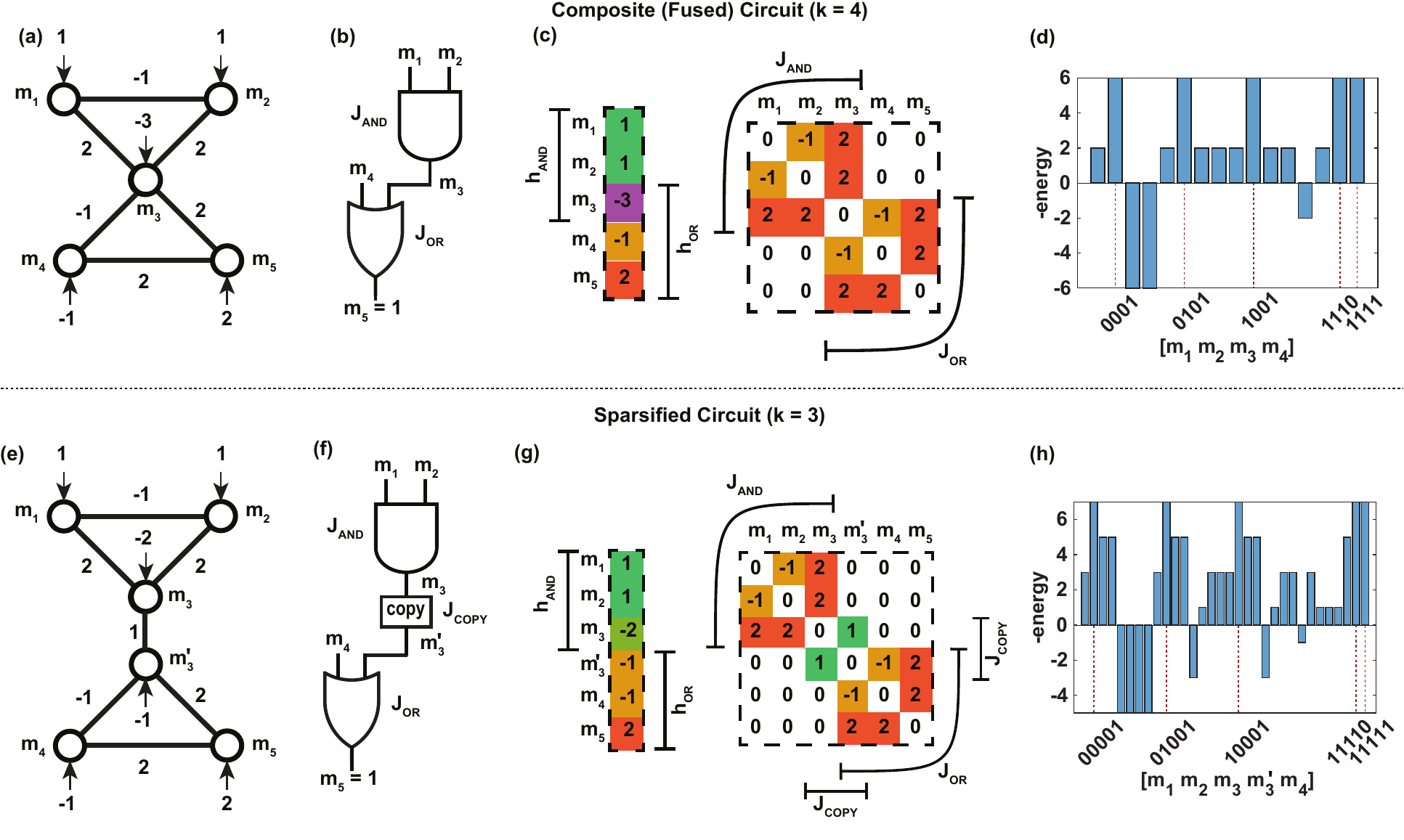}
    \caption{Fusion and sparsification techniques for probabilistic gates. (a)-(d) A composite p-circuit built using an AND gate and an OR gate with corresponding $J$ and $h$ matrices and energies of states. Unlike the other examples, $m_5$ is clamped to 1 in this example. (e)-(h) A sparse p-circuit built by splitting $m_3$ into two nodes and inserting a copy gate between $m_3$ and $m_3'$. The energy plot shows the same ground states, showing that these two circuits are equivalent.}
    \label{fig:circuit}
\end{figure*}

Fig.~\ref{fig:circuit}a-d further illustrates how to compose an invertible logic p-circuit using these basic logic gates. The composite circuit combines the $[3\times3]$ $J$ matrices of the AND and OR gates which become a $[5\times5]$ matrix after fusion of the common node. Similarly, it combines the $[3\times1]$ $h$ matrices of the AND and the OR gates which become a $[5\times1]$ $h$ matrix after fusion. The energy plot reveals that the ground states $[m_1~m_2~m_3~m_4]$ = $\{0001, 0101, 1001, 1110, 1111\}_{\rm bin}$  agree with the truth table of the composite circuit when $m_5$ = 1. The circuit has a maximum number of neighbours, $k = 4$. A sparser version of the circuit with $k = 3$ can be obtained by splitting $m_3$ into two nodes and inserting a copy gate between $m_3$ and $m_3'$ (Fig.~\ref{fig:circuit}e-h). The ground states are $[m_1~m_2~m_3~m_3'~m_4]$ = $\{00001, 01001, 10001, 11110, 11111\}_{\rm bin}$ when $m_5$ = 1. We present a mathematical justification in Supplementary Section~\ref{sec:fs} showing the equivalence between these two circuits. 

\subsection{Fusion and Sparsification}
\label{sec:fs}

In this section, our objective is to define two graph modification techniques we call fusion and sparsification. Fusion refers to combining multiple nodes (p-bits) to a single node. Sparsification refers to splitting a single node into multiple nodes to decrease the vertex degree (number of neighbors) of a given node. We define $k$ as the maximum number of neighbors per node in a graph. $k=\infty$ represents the original problem (fused) without sparsification. In this section, we give a mathematical proof of the equivalence between a fused and a sparisified graph.

\subsubsection{Mathematical equivalence of fused and sparsified graphs} 
\label{sec:equivalence}
Here, we would like to establish that as the annealing parameter $\beta$ of Eq.~\eqref{eq:pbit} (main paper) is increased ($\beta\rightarrow\infty$ at the end of an annealing schedule), the sparsified and the fused circuits have the same ground state for a given optimization problem. The connection between the sparsification and the fusion also allows a natural way of composing p-circuits which we elaborate next. Suppose there are two subcircuits we are trying to connect in which $m_A$ corresponds to the node from subcircuit A and $m_B$ corresponds to the node from subcircuit B and that $m_A$ and $m_B$ need to be tied together. Both these systems are described by their respective energies (ignoring biases without loss of generality):
\begin{equation}
    E_A =-\left(\sum_{\substack{i\neq A\\ 
                  i<j}} J_{ij} m_i m_j + \sum_{j} J_{Aj} m_j m_A \right)
                  \label{eq:enA}
         \end{equation}
         \begin{equation}
             E_B =-\left(\sum_{\substack{i\neq B\\ 
                  i<j}} J'_{ij} m_i m_j + \sum_{j} J'_{Bj} m_j m_B\right)
                  \label{eq:enB}
         \end{equation}
where we separated the energy terms corresponding to $m_A$ and $m_B$ from the rest of the subcircuits. If $m_A$ and $m_B$ are to be connected as a common node between these subcircuits, as in ordinary digital circuits, a positive interaction parameter (ferromagnetic, $J_T >0$) can be used to connect them such that the total energy of the composed system is given as:
\begin{equation}
    E =E_A + E_B - J_T m_A m_B  
    \label{eq:composite} 
\end{equation}
This situation corresponds to the sparsified network where the interaction parameter $J_T$ corresponds to the COPY gate which ties the same logical value to $m_A$ and $m_B$. 

We continue the analysis with a physical observation. As the temperature is lowered ($\beta\rightarrow\infty$), $m_A$ and $m_B$ cannot differ in their states due to the large energy penalty incurred by $\beta J_T$, or mathematically, $P(m_A\neq m_B)=\exp(-\beta J_T) \rightarrow 0 $, independent of the specific value of $J_T$. This allows, in the bipolar notation where $m_A, m_B \in \pm 1 $, the following trick: $m_A=m_B$ and $m_A^2=1$. This means that the last term in Eq.~\eqref{eq:composite} becomes a constant and drops out of the final Boltzmann probabilities since any constant term in the energy cancels out:
\begin{equation}
    P(m_1,\ldots,m_N) = \frac{1}{Z} \exp(\beta J_T) \exp[-\beta (E_A+E_B)]
\end{equation}
where 
 \begin{equation}
    Z = \sum \exp(-\beta E) = \exp(\beta J_T) \sum \exp -\beta (E_A+E_B)
\end{equation}

This analysis indicates that for an annealed system ($\beta\rightarrow \infty$), irrespective of the strength of the coupling parameter $J_T$, there is no difference in the final probabilities between subcircuits A and B. For example, subcircuit A can be the fused circuit  shown in Fig.~\ref{fig:circuit}a-d and  subcircuit B can be the sparsified circuit shown in Fig.~\ref{fig:circuit}e-h. This analysis is similar to the behavior of the many replicas collapsing to a single qubit in the Suzuki-Trotter transformation, enabling a mapping between the thermodynamics of a many-body quantum system and a probabilistic system \cite{suzuki1976relationship}. 

Going back to Eq.~\eqref{eq:enA}-\eqref{eq:enB}, and substituting $m_A=m_B$ and calculating the input to the node $m_A$ by $I_A=-\partial E/\partial m_A$: 
\begin{equation}
    I_A = \sum J_{Aj} mj + \sum J'_{Bj} m_j
    \label{eq:composing}
\end{equation}
Eq.~\eqref{eq:composing} shows the mathematical justification of adding the columns of fused nodes together, as shown in the composite circuit of  Fig.~\ref{fig:circuit}a-d. Note that the rows of fused nodes also need to be added together to ensure the symmetry of the J matrix.

\subsubsection{Fused circuit}
\label{sec:fused_circuit}
For  the $n$-bit factorizer circuit in Fig.~\ref{fig:fig1}a, we have $2m$ p-bits from the input bits of the AND gates. The output p-bits of the AND gates get fused with the corresponding input p-bits of the FAs, except the direct output of the first AND gate that represents $\rm S_0$. It will be represented by a single p-bit. For the FAs, the first row has $4m+1$ p-bits instead of $5m$, since all the neighbor FAs have the $\rm c_{out}$ and $\rm c_{in}$ fused together. For the other rows of the FAs, we have $3m+1$ p-bits per row. This is because one of the input bits comes from the previous row. The number of p-bits in a fused $n$-bit factorizer p-circuit can be generalized as
\begin{equation}
\begin{aligned}
\operatorname{\textit{N}_{fact}^{(fused)}} &=2 m+(4 m+1)+(3 m+1)(m-2)+1 \\
&=3 m^{2}+m
\end{aligned}
\end{equation}
where $m = \frac{n}{2}$.\vspace{3mm}

Likewise, an invertible logic 3SAT solver circuit (a special case of Fig.~\ref{fig:fig1}b with exactly 3 variables per clause and needs only 2 rows of OR gates) can be composed using fusion. Same input variable routed to multiple places is represented by a single p-bit instead of multiple p-bits. This way, we have exactly one p-bit for each input variable. The output p-bits of the OR gates in the first row get fused with the corresponding input p-bits of the OR gates in the second row. To encode this, we need exactly one p-bit for each clause. Finally, all the output p-bits of the OR gates in the second row are fused together and represented by a single p-bit clamped to 1. The number of p-bits in a fused 3SAT p-circuit can be generalized as

\begin{equation}
\operatorname{\textit{N}_{3SAT}^{(fused)}} = c + v + 1
\end{equation}
where, $c$ = number of clauses, and $v$ = number of input variables.

The fused p-circuit is software-friendly since it keeps the state space smaller, however, it introduces a fan-out issue in the sIM due to having too many neighbors for some p-bits. It also slows down the clock speed as discussed in Section \ref{sec: sparsification} in the main manuscript.

\subsubsection{Sparsified circuit}
\label{sec:sparse_circuit}

In a sparsified p-circuit, we do not fuse the p-bits when it exceeds a predefined maximum number of neighbors, $k$ for any p-bit. Here, we demonstrate two different ways of sparsifying a graph, one with an example of the integer factorization and the other with an example of the 3SAT solver.

In the factorizer p-circuit, we add a series of p-bits to the same-signal input bits using copy gates. Having set a maximum number of neighbors, $k$ for each p-bit, we add a p-bit for every $(k-1)$ input bits to be connected. If the total number of p-bits added is more than 1, the process is started again until only one p-bit is added, which represents the actual input p-bit. 

If no p-bits are fused, the $n$-bit factorizer circuit in Fig.~\ref{fig:fig1}a has $3m^2$ p-bits for the $m^2$ AND gates and $5m(m-1)$ p-bits for the $m(m-1)$ FAs. The number of p-bits for a sparsified $n$-bit factorizer p-circuit with a maximum number of neighbors, $k$ per p-bit includes additional p-bits for the copy gates and can be generalized as
\begin{equation}
\begin{aligned}
\operatorname{\textit{N}_{fact}^{(sparse)}}&= 3m^{2} + 5m(m-1) + 2m f(m,k)\\
&=8m^{2} - 5m + 2m f(m,k)
\end{aligned}
\end{equation}
where
\begin{align*}
m &= \frac{n}{2}\\
f(m, k) &=\sum_{i=1}^{\lceil\log _{k-1} m\rceil} a_{i} \\
a_{0} &=m \\
a_{i} &=\bigg\lceil{\frac{a_{i-1}}{k-1}}\bigg\rceil
\end{align*}

In this work, we set $k = 5$ for integer factorization and the expression can be approximated as
\begin{equation}
\begin{aligned}
\operatorname{\textit{N}_{fact}^{(sparse)}}&\approx 8 m^{2}-5 m+2 m\bigg\lceil\frac{m-1}{3}\bigg\rceil \\
&\approx 8 m^{2}-5 m+2 m \frac{m-1}{3} \\
&=\frac{26}{3} m^{2}-\frac{17}{3} m
\end{aligned}
\end{equation}

For the 3SAT problem, instead of fusing, we connect the same-signal input bits by inserting a copy gate between every two input p-bits. This way, the input p-bits get correlated and also avoid additional neighbors. The output p-bits of the OR gates in the first row get fused with the corresponding input p-bits of the OR gates in the second row as before since it does not cost any fan-out issue. Finally, the output p-bits of the OR gates in the second row are fused in pairs and clamped to 1. Each clause has 2 OR gates and thus 5 p-bits since the middle p-bits get fused. However, since the output p-bits also get fused in pairs, we have 1 p-bit less for every 2 clauses. The final sparsified 3SAT solver p-circuit has a maximum number of neighbors  $k = 4$ only and the number of p-bits can be generalized as ($c$ = number of clauses)
\vspace{-15pt} 

\begin{equation}
\operatorname{\textit{N}_{3SAT}^{(sparse)}}  = \lceil5c - \frac{1}{2}c\rceil = \lceil\frac{9}{2}c\rceil
\end{equation}

The sparsified p-circuit is hardware-friendly as it limits fan-out and allows fast clocks with small adder delay in the sIM.

\subsubsection{Graph density} 
\label{sec:density}

In Section~\ref{sec: sparsification} of the main manuscript, we have discussed how the sparse (less dense) graphs reduce adder delays by limiting the maximum number of neighbors, $k$ in a graph. Here, we add an analytical expression for the maximum graph density and show how sparsity helps to scale the proposed architecture.

For a graph with $k$ regular neighbors and $|V|$ nodes (vertices), the total number of edges ($|E_{\sf k\mbox{-}regular}|$) is
\begin{equation}
\begin{aligned}
|E_{\sf k\mbox{-}regular}|&= \displaystyle \frac{k|V|}{2}\\
\end{aligned}
\label{eq:E_regular}
\end{equation}

For a graph with all-to-all connections and $|V|$ nodes, the total number of edges ($|E_{\sf all\mbox{-}to\mbox{-}all}|$) is
\begin{equation}
\begin{aligned}
    |E_{\sf all\mbox{-}to\mbox{-}all}|&=  \displaystyle\frac{|V|^2-|V|}{2}\\
\end{aligned}
\label{eq:E_alltoall}
\end{equation}

Hence, a graph with a maximum of $k$ neighbors will have maximum graph density ($\rho_{\sf max})$:
\begin{equation}
\begin{aligned}
\rho_{\sf max} = \displaystyle\frac{|E_{\sf k\mbox{-}regular}|}{|E_{\sf all\mbox{-}to\mbox{-}all}|} = \displaystyle\frac{k}{|V|-1} \approx \displaystyle\frac{k}{|V|}\\
\label{eq:rho_max}
\end{aligned}
\end{equation}

\begin{figure*}[t!]
    \centering
    \includegraphics[width=0.75 \textwidth]{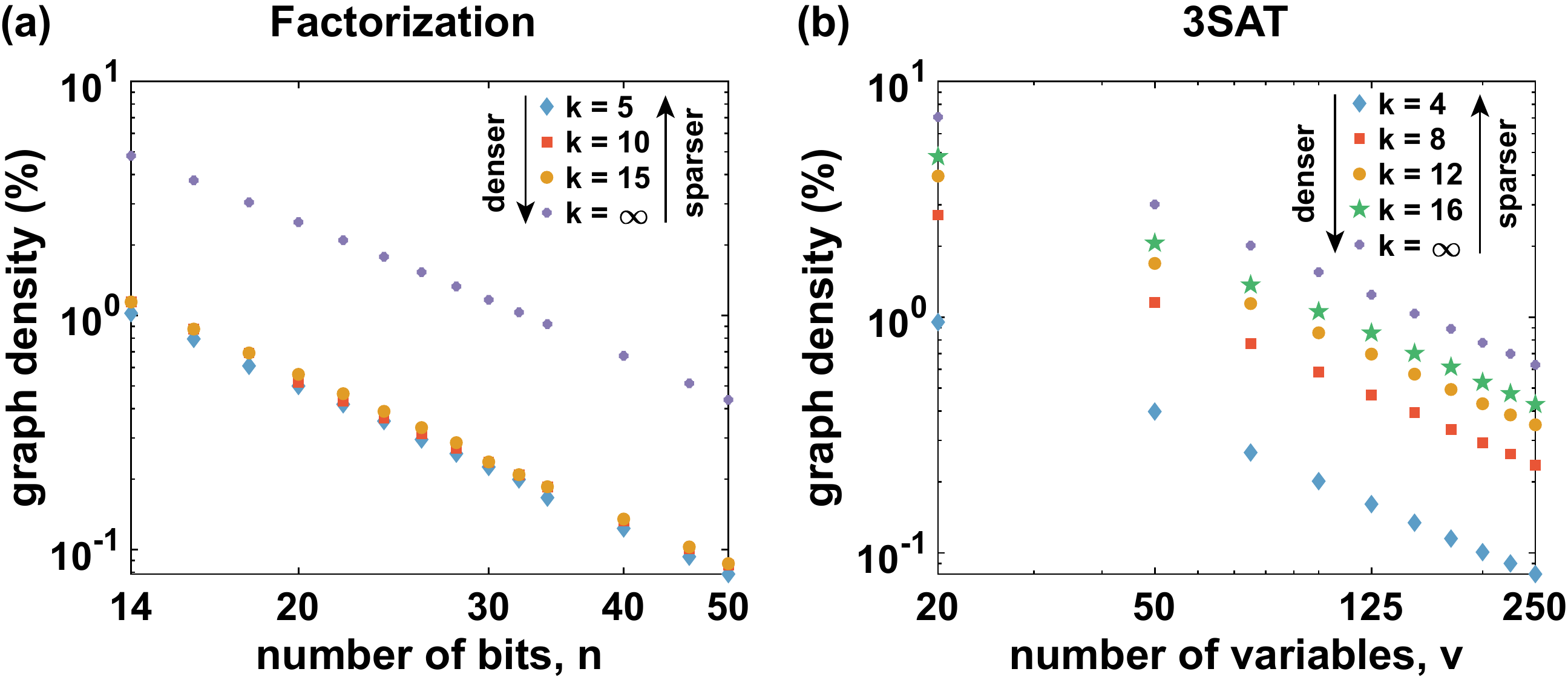}
    \caption{Graph density as a function of problem size at different $k$ values for (a) factorization and (b) 3SAT problem. $k$ is defined as the maximum number of neighbors after sparsification. $k=\infty$ represents the original problem without sparsification.}
    \label{fig:density}
\end{figure*}

As Fig.~\ref{fig:density} shows, for both the integer factorization and 3SAT instances, the graph density (Eq.~\eqref{eq:rho}) as a function of problem size progressively \textit{decreases} even without sparsification ($k=\infty$), hence they can be efficiently represented in a sparse, scalable hardware. 

\begin{figure*}[t!]
    \centering
    \includegraphics[width=0.75 \textwidth]{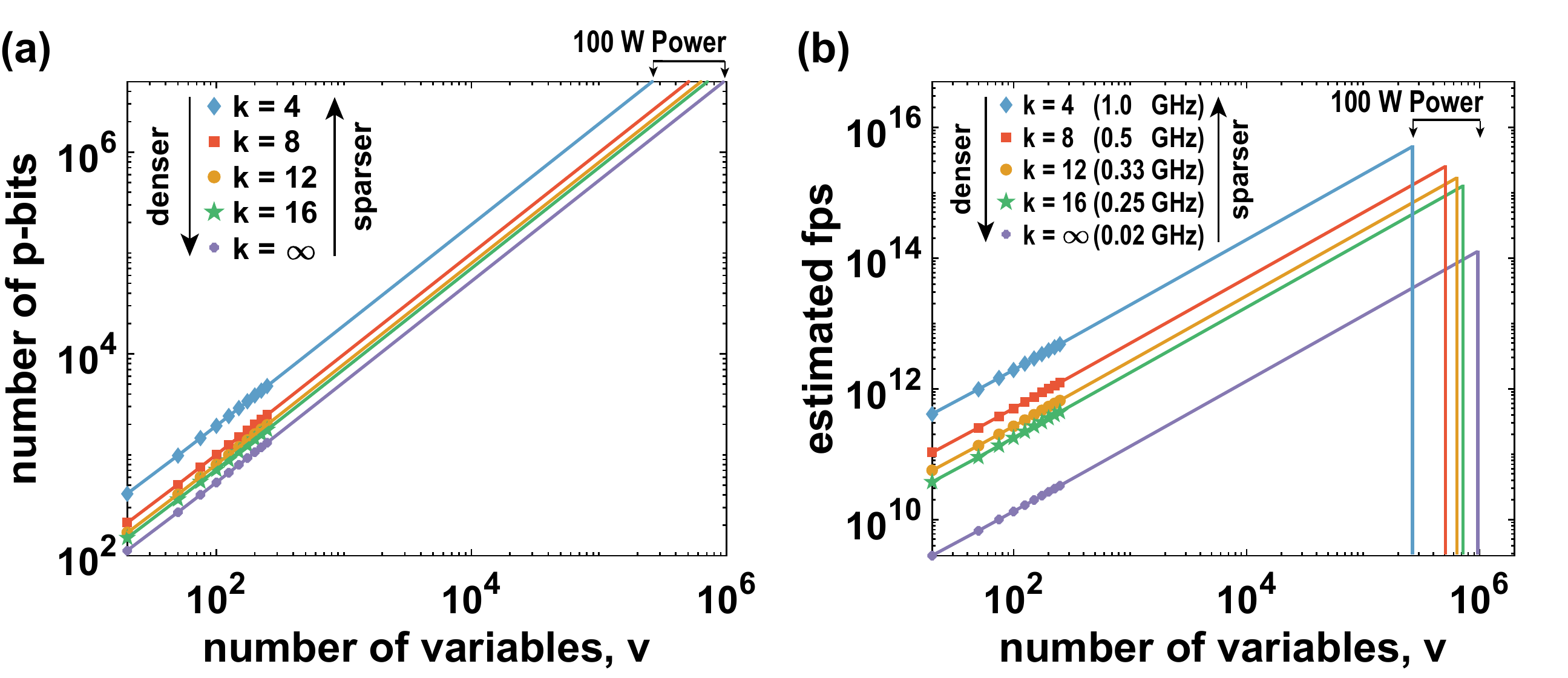}
    \caption{Performance Projections. We consider projections up to a million p-bits for the 3SAT problem, assuming 20 $\mu$W per p-bit inspired by detailed device simulations \cite{hassan2021quant}. (a) Number of p-bits as a function of problem size at different $k$ values. (b) estimated fps as a function of problem size at different $k$ values operating at different clock frequencies (inversely depending on $k$). For a given power budget of 100 W, the sparsest representation allows the fastest fps, however, for further scaling, denser graphs can be used with lower fps. $k$ is defined as the maximum number of neighbors after sparsification. $k=\infty$ represents the original problem without sparsification.}
    \label{fig:scalability}
\end{figure*}

\subsection{Scalability analysis}
\label{sec:scalability}
The proposed architecture has two main parts: p-bits and interconnections. Both p-bits and interconnections scale linearly in terms of resources. Each p-bit uses a fixed amount of resources (e.g., LUT, LFSR) which grows linearly with increasing number of p-bits [O($N$)]. Similarly, the interconnections also grow linearly [O($N$)], since we limit the maximum neighbors for a p-bit to a fixed number,  $k$ = 4, $k$ = 8, etc.

\subsubsection{Trade-off between resources and performance}
\label{sec:tradeoff}
The optimum sparsity of a problem depends on a trade-off between resources (e.g, number of p-bits, interconnects) and performance (e.g., flips per second). The number of p-bits and the number of interconnects are limited by a given power and area budget. For example, Fig. \ref{fig:scalability} shows nanodevice (Magnetic Tunnel Junction) based projections for the 3SAT problem. Using MRAM technology we project that up to a million p-bits can be integrated within a power budget of 100 W since each p-bit dissipates around 20 $\mu$W based on detailed device simulations \cite{hassan2021quant}. For further scaling, using denser graphs that can accommodate larger problems is possible, however, this approach shows diminishing returns beyond a point, where increasing the graph density does not help if the original problem (e.g., Boolean SAT) is already sparse (Fig. \ref{fig:scalability}a). 

While the sparsest representations allow the fastest flips per second (Fig. \ref{fig:scalability}b), they lead to an increase in the number of p-bits. The operating clock frequency decreases linearly with the maximum number of neighbors, $k$ since we assume the adder delay increases linearly as a function of $k$.

\begin{table}[t!]
    \centering
    \begin{tabular}{|c|c|c|c|c|c|}
        \hline
        \multirow{2}*{\bf Bits} & \multirow{2}*{\bf Original Graph} &{\bf Sparsification} &{\bf MGE (Chimera Graph)} &{\bf MGE (King's Graph)} &{\bf MGE (Grid Graph)} \\
        & & \bf[This Work] & \bf Fixed Spins $\approx$ 50000 & \bf Fixed Spins $\approx$ 50000 &  \bf Fixed Spins  $\approx$ 50000\\
        \hline 
        \hline 
        14 & 154 spins & 2128 spins & 1162 out of 50000 spins & 1690 out of 50000 spins & Fails\\
        16 & 200 spins & 2128 spins & 1465 out of 50000 spins & 2734 out of 50000 spins & Fails\\
        18 & 252 spins & 2128 spins & 1946 out of 50000 spins & 3552 out of 50000 spins & Fails \\
        20 & 310 spins & 2128 spins & 2888 out of 50000 spins & Fails & Fails \\ 
        22 & 374 spins & 2128 spins & 3108 out of 50000 spins & Fails & Fails \\ 
        24 & 444 spins & 2128 spins & 4766 out of 50000 spins & Fails & Fails \\
        26 & 520 spins & 2128 spins & 5786 out of 50000 spins & Fails & Fails \\
        28 & 602 spins & 2128 spins & 6017 out of 50000 spins & Fails & Fails \\
        30 & 690 spins & 2128 spins & 8896 out of 50000 spins & Fails & Fails \\
        32 & 784 spins & 2128 spins & 10320 out of 50000 spins & Fails & Fails \\
                        
        \hline
    \end{tabular}
    \caption{Minor graph embedding (MGE) vs. invertible Boolean logic embedding for the integer factorization problem. We assume a fixed $\approx$ 50000-spin Chimera, King's and square grid topologies as target graphs for MGE. For the sparse Ising Machine, a fixed 2128-spin hardware can factor all integers up to 32-bits. }
    \label{tab:MGE_fact}
\end{table}

\subsubsection{Invertible Boolean logic  vs. minor graph embedding}
\label{sec:MGE} 
Here, we report an illustrative comparison between invertible Boolean logic vs. minor graph embedding (MGE) applied to the integer factorization problem. For MGE, we use D-wave's minor-miner program \cite{cai2014practical} and assume that a fixed hardware with $\approx$ 50000 spins in Chimera, King's and square grid graph topologies needs to embed an original graph to factor different sizes of semiprimes, up to 32-bits. 

For MGE, Table~\ref{tab:MGE_fact} shows that the Chimera graph requires $\approx$ 10000 spins to encode the 32-bit factorizer. The King's graph fails to encode beyond 18-bits and the grid graph always seems to fail. For the sparse Ising Machine, on the other hand, a 32-bit invertible multiplier can factor any number up to 32-bits. Only a sparsified graph with $k=4$ having 2128 spins that can factor 32-bits is necessary and sufficient at all sizes.
     
\subsection{Time to solution for exact factorization}
\label{sec:TTS_100}
\begin{figure*}[t!]
    \centering
    \includegraphics[width=0.72 \textwidth]{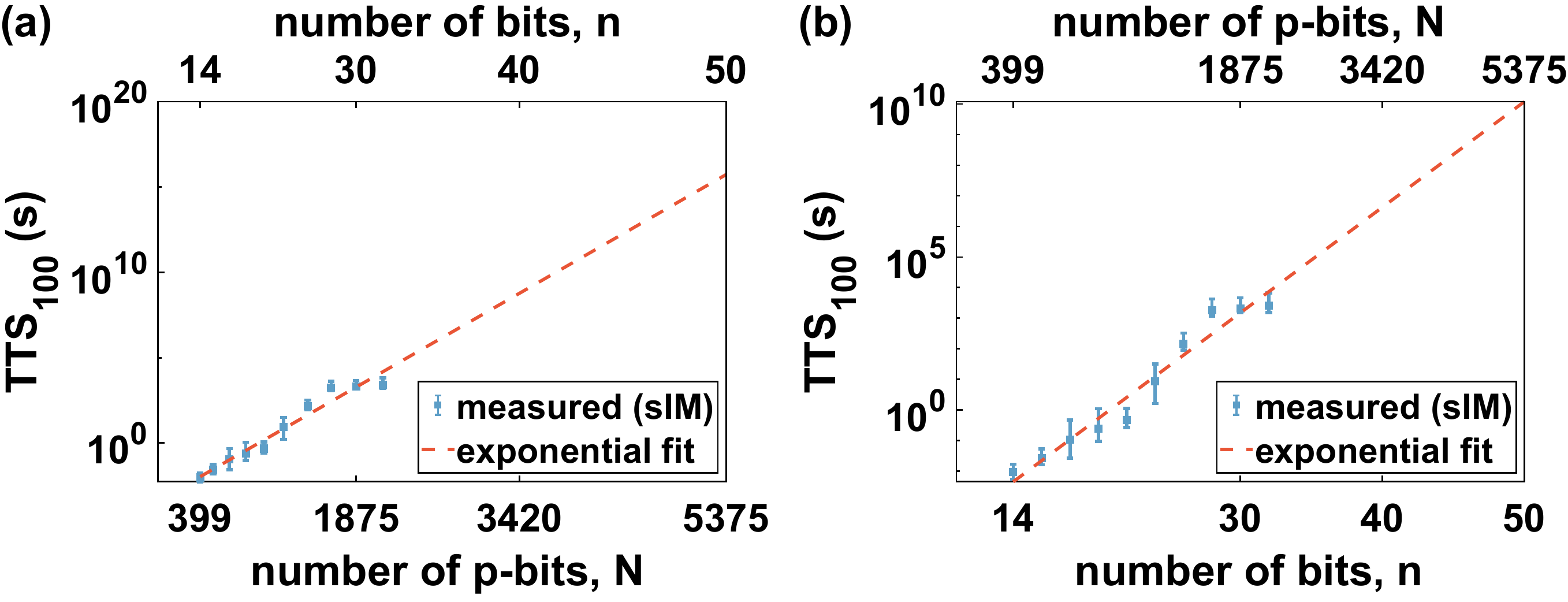}
    \caption{Time to solution for exact factorization ($\rm TTS_{100}$) of 14-bit to 32-bit semiprime numbers in the sIM and an exponential fit with respect to the (a) number of p-bits up to 5375 p-bits and (b) number of bits up to 50-bits.}
    \label{fig:TTS_100}
\end{figure*}

We have reported exact factorization up to 32-bit semiprime numbers in Section~\ref{sec:exact_fact}. Here, we report the time to find the exact factors. An exponential fit with respect to the number of p-bits up to 5375 p-bits shows the difficulty in factoring numbers larger than 32-bit with the current annealing schedule (Fig.~\ref{fig:TTS_100}a). We describe this plot with the following equation:
\begin{equation}
\label{eq:tau1}
t=t_{0} \exp \left(\frac{N}{\tau}\right)
\end{equation}
where $N$ = number of p-bits, $t$ = $\rm TTS$, $t_0$ = pre-exponential factor, $\tau$ = time constant. From the fitted plot, we obtain  $t_0 = 10^{-3.39} s$ and $\tau = 122.13$. 

Another exponential fit with respect to the number of bits up to 50-bits is shown in Fig.~\ref{fig:TTS_100}b. We describe this plot with the following equation:
\begin{equation}
\label{eq:tau2}
t=t_{0} \exp \left(\frac{n}{\tau}\right)
\end{equation}
where $n$ = number of bits. 
From the fitted plot, we obtain  $t_0 = 10^{-7.17} s$ and $\tau = 1.26$. 

\subsection{Error models for inexact Gibbs sampling}
\label{sec:mask_model}
In Section~\ref{sec: overclocking} of the main manuscript, we show how moderate overclocking leads to a decrease in the time to solution in two different problems. In order to analyze this phenomenon, we introduce two error models for inexact Gibbs sampling and apply them to systematically study a 5 p-bit full adder (FA) circuit. However, the conclusions and limits we obtain are generally applicable. The $J$ and $h$ matrices we used for the full adder are the following:

\begin{align}
\label{eq:fulladder}
J_{FA}=
\begin{pmatrix}
0 & -1 & -1 & +1 & +2 \\
-1 & 0 & -1 & +1 & +2 \\
-1 & -1 & 0 & +1 & +2 \\
+1 & +1 & +1 & 0 & -2 \\
+2 & +2 & +2 & -2 & 0
\end{pmatrix}
&&
h_{FA}=
\begin{pmatrix}
0 & 0 & 0 & 0 & 0
\end{pmatrix}
\end{align}

The two models are both based on the same principle: the $J$ matrix is multiplied by an error mask matrix whose elements establish whether a neighbor connection is failing or not. If no nodes are failing, both mask models are a matrix of 1s. 

In the first model we introduce, which we will refer to as the single mask model from here on, uses a mask of $+1$s and $-1$s to represent functioning and failing connections, respectively. For example, if a connection from node $j$ to node $i$ is failing due to overclocking,  $M^S_{ij}$ will be $-$1.   Thus, Eq.~\eqref{eq:syn} in the main text will be replaced by

\begin{equation}
\label{eq:onemaskmodel}
I_i = \sum_{j}M^S_{ij}J_{ij} m_{j} + h_{i}
\end{equation}
where $M^S$ is the error mask for the single mask model. In the case considered, since $J_{FA}$ is a $5\times5$ matrix and since the main diagonal elements cannot fail as a connection, the maximum number of errors is 20.

The second model, which we will refer to as the double mask model from here on, uses two complementary masks of $1$s and $0$s. In this case, Eq.~\eqref{eq:syn} will be replaced by

\begin{equation}
\label{eq:twomasksmodel}
I_i = \sum_{j}M^D_{ij}J_{ij} m_{j}^{\rm new} + \sum_{j}(1-M^D_{ij})J_{ij} m_{j}^{\rm old} + h_{i}
\end{equation}
where $m_j^{\rm new}$ and  $m_j^{\rm old}$ represent the updated and the non-updated values, respectively. 

\begin{figure*}[t!]
    \centering
    \includegraphics[width=1.00 \textwidth]{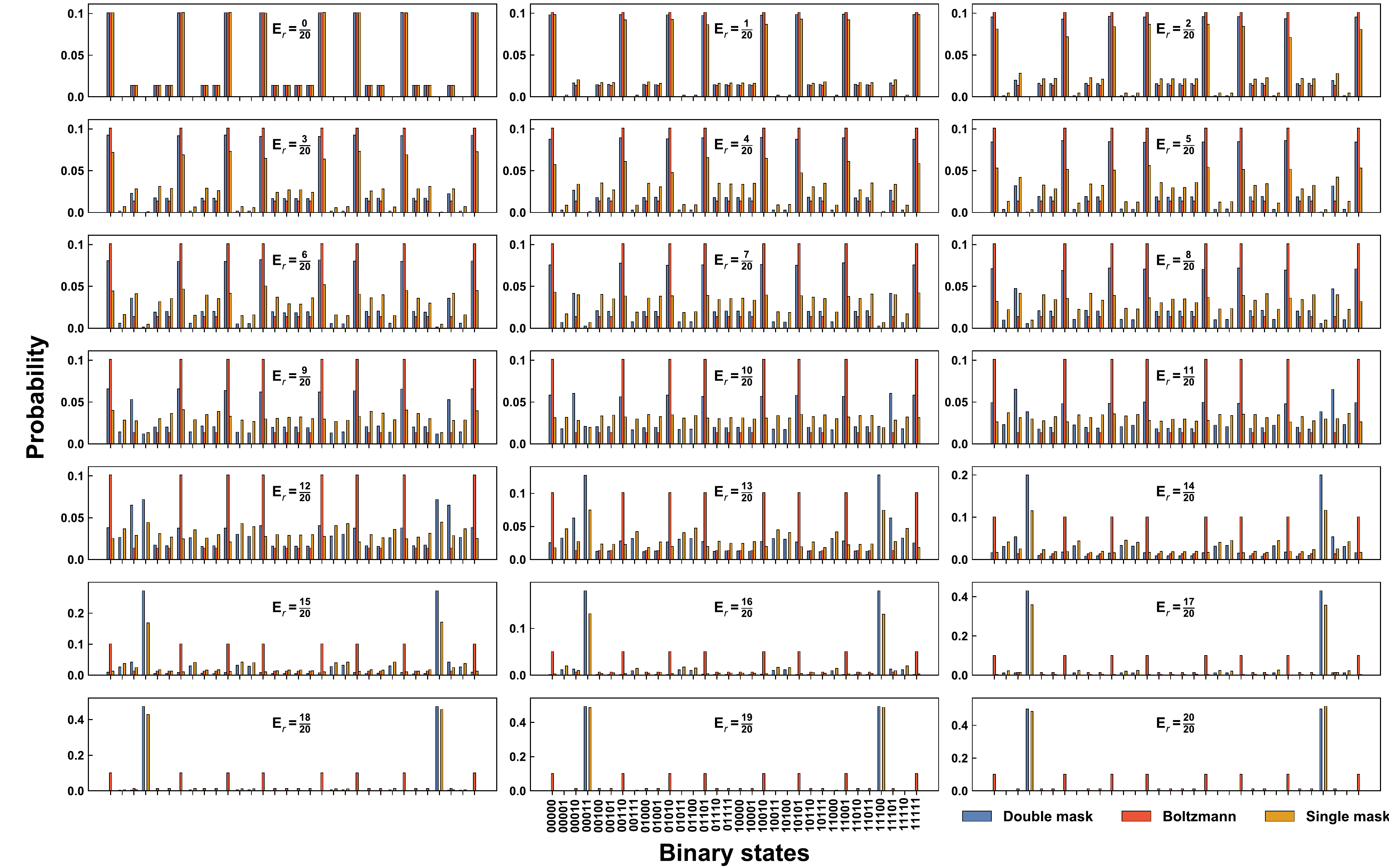}
    \caption{Probability distributions of the single and double mask models vs. the exact Boltzmann distribution as a function of the fraction of errors $E_r$ for a 5 p-bit full adder. Each plot is obtained by averaging over 400 random masks, each sampled $2\times10^4$ times with $\beta=1$. Both models show relatively low deviations from the exact distribution at low values of $E_r$ before dramatically worsening and converging to a parallel update distribution.}
    \vspace{-10pt}
    \label{fig:distribution_error}
\end{figure*}

\begin{figure}[h]
    \centering
    \includegraphics[width=0.48 \textwidth]{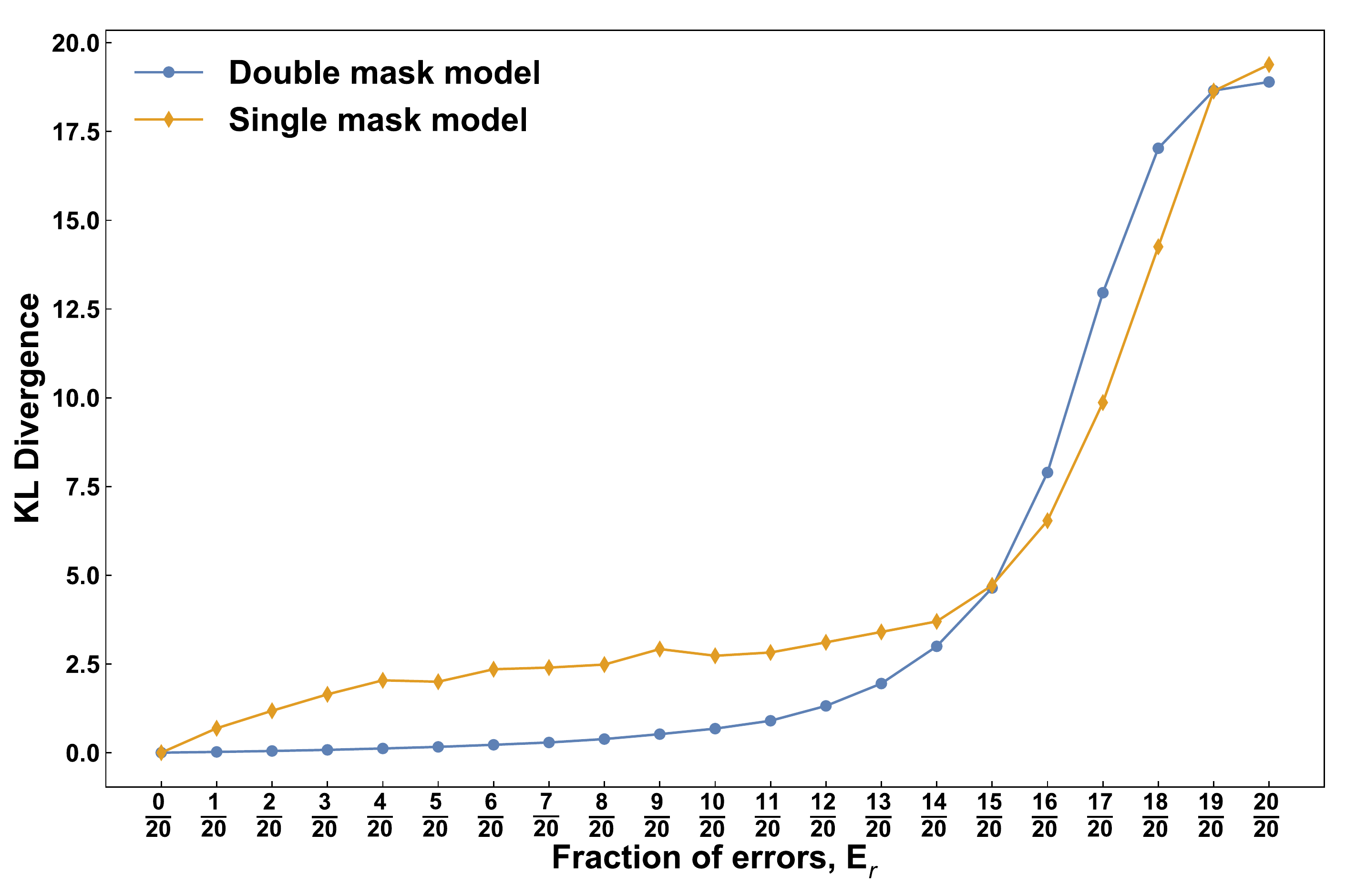}
    \caption{Kullback–Leibler (KL) divergence of the distributions of the single and double mask models from the exact Boltzmann distribution as a function of the fraction of errors $E_r$. Both models exhibit (albeit with different degrees of accuracy to the real distribution) a flatline behavior with moderate values of $E_r$ and a sharp increase in divergence when $E_r$ reaches a certain threshold.}
    \label{fig:KL}
\end{figure}

To test these models, we performed a systematic study where we considered every possible fraction of errors (defined as $E_r$) in the mask matrices (from $0/20$ to $20/20$). We simulated 400 random masks for each $E_r$ taking $2\times10^4$ samples per mask. In Fig.~\ref{fig:distribution_error}, the average distributions resulting from all masks for each value of $E_r$ are  compared with the exact Boltzmann distribution. Both models exhibit qualitatively similar behavior. For a moderate number of errors the Kullback–Leibler (KL)  divergence from the exact distribution does not increase significantly, but beyond a certain point errors diverge (Fig.~\ref{fig:KL}). Eventually both models approach a distribution described by a parallelly updating network  \cite{aarts1989simulated} whose steady-state is defined by the following equation:

\begin{equation}
\label{eq:parallel_analytical}
p_k = p(m_1^{(k)}, m_2^{(k)}, \cdots, m_N^{(k)}) = \frac{1}{Z} \prod_{i=1}^{N}  \ \mathrm{cosh}\left(\beta\left[\sum_{j} J_{ij} m_{j}^{(k)} + h_{i}\right]\right) \mathrm{exp}\left(\beta \ m_i^{(k)} h_i\right)   
\end{equation}
where $k$ represents all possible states of $\mathbf{m}$ from $1, 2, \ldots 2^N$ and $Z$ is a normalization constant ensuring probabilities add to 1. 

The distribution defined by Eq.~\eqref{eq:parallel_analytical} quantitatively describes the steady-state distribution observed in Fig.~\ref{fig:distribution_error} as $E_r$ approaches 1, at which point the network updates are parallel. 

\null
\vfill

\end{document}